\documentclass[aps,superscriptaddress,eqsecnum,nofootinbib,showpacs,preprintnumbers]{revtex4-2}
\usepackage{amsmath}
\usepackage{amssymb}
\usepackage{amsfonts,amsthm,bm}
\usepackage{color,xcolor}
\usepackage[greek,english]{babel}
\usepackage{graphicx,epsfig}
\graphicspath{{figures/}{./}}
\usepackage{float}
\usepackage{subfigure}
\usepackage{tikz}
\usepackage[pdfencoding=auto,colorlinks=true,
allcolors=blue]{hyperref}

\newcommand{\abs}[1]{\left| #1 \right|}

\newcommand{\be}{\begin{eqnarray}}
\newcommand{\ee}{\end{eqnarray}}
\newcommand{\bea}{\begin{eqnarray}}

\newcommand{\eea}{\end{eqnarray}}

\def\D{\Delta}

\def\z{\zeta}

\def\Ref{\ref}

\DeclareMathOperator{\arctanh}{arctanh}
\newcommand{\beq}{\begin{equation}}
\newcommand{\eeq}{\end{equation}}
\newcommand{\bseq}{\begin{subequations}}
	\newcommand{\eseq}{\end{subequations}}

\begin{document}
	
	\title{Formation of Bound States of scalar fields in AdS-asymptotic Wormholes}
	
	\author{Nikos Chatzifotis}
	\email{chatzifotisn@gmail.com}
	\affiliation{Physics Department, National Technical University of Athens, 15780 Zografou Campus, Athens, Greece}
	
	\author{George Koutsoumbas}
	\email{kutsubas@central.ntua.gr}
	\affiliation{Physics Department, National Technical University of Athens, 15780 Zografou Campus, Athens, Greece}
	
	\author{Eleftherios Papantonopoulos}
	\email{lpapa@central.ntua.gr} \affiliation{Physics Department, National Technical University of Athens, 15780 Zografou Campus, Athens, Greece}

	
	\begin{abstract}
	We use the  Wentzel-Kramers-Brillouin (WKB) approximation to study the formation and propagation of bound states in the vicinity of a wormhole in the non-minimal derivative coupling theory of gravity.  The wormhole throat connects two Anti-de Sitter spacetimes. We show that when the scalar field lies in high orbital states, the corresponding potential has potential barriers that block the passage of a classical field. We investigate the behaviour of the bound states trapped in the potential wells and provide the flow between the two AdS regions.
	\end{abstract}

	\maketitle
	
	\section{Introduction}

Wormholes in General Relativity (GR) are solutions of Einstein equations that connect different parts of the Universe or two different Universes.
 The concept of wormhole was introduced in the  pioneering articles of Misner and Wheeler \cite{MisWheel} and Wheeler \cite{Wheel}.
 Lorentzian wormholes in  GR were  studied by Morris and Thorne \cite{Morris} where a static spherically symmetric metric was introduced and conditions for traversable wormholes were found. However, a condition on the wormhole throat leads to the violation of the null energy condition (NEC). A matter distribution of exotic or phantom matter allows in GR the formation of traversable wormhole geometries. This type of matter has been  discussed in cosmological contexts \cite{phantom}, for possible  observational settings. There have been many efforts to build a wormhole with ordinary matter satisfying the NEC. In \cite{Visser}  the construction of thin-shell wormholes was  studied, where the supporting matter is concentrated   on the wormhole throat. In \cite{Mehdizadeh:2015dta} it was shown that in theories were higher order terms in curvature are present it is possible to build  thin-shell wormholes supported by ordinary matter.
 Recently there are many studies of wormhole solutions in modified gravity theories like Brans-Dicke theory \cite{Brans}, $f({\sf R})$ gravity \cite{FR},
 Einstein-Gauss-Bonnet theory~\cite{GB}, Einstein-Cartan theory and general scalar-tensor theories \cite{Cartan}.

The simplest  and very well studied modifications of GR are the scalar-tensor theories \cite{Fujii:2003pa}. The presence of a scalar field coupled to gravity has important implications in local and global solutions in these theories.  The Horndeski Lagrangian
\cite{Horndeski:1974wa} provides one of the best studied    scalar-tensor theories. The reason is that the Horndeski theories
lead to second-order field equations, they give rise to
consistent theories without ghost instabilities \cite{Ostrogradsky:1850fid, Woodard:2006nt, Woodard:2015zca, Deffayet:2011gz} and they preserve a classical
Galilean symmetry \cite{Nicolis:2008in,Deffayet:2009wt}. The Horndeski theory has been studied in short and large distances.
In particular a subclass of Horndeski theories was studied in which the scalar field  is kinetically coupled to the Einstein tensor.
Then black hole solutions were found \cite{Kolyvaris:2011fk,Rinaldi:2012vy,Kolyvaris:2013zfa,Babichev:2013cya,Charmousis:2014zaa}, known as  Galilean black holes, and also wormhole
geometries were generated \cite{Korolev:2014hwa,Korolev:2020ohi,Korolev:2020yyy}.
At large distances  the presence of the derivative coupling acts as a friction term in the inflationary period of the cosmological
evolution  \cite{Amendola:1993uh,Sushkov:2009hk,Germani:2010hd,Saridakis:2010mf,Huang:2014awa,Yang:2015pga,Koutsoumbas:2013boa}. This derivative coupling
introduces a mass scale in the theory which can be constrained, at large distances, by the recent results on gravitational waves (GWs) \cite{Gong:2017kim}.

The effects of the Galileon black holes were studied in \cite{Koutsoumbas:2018gbd}. Considering a test wave in the vicinity of a Galileon black hole it was shown that a Regge-Wheeler potential  arose and the formation and the behaviour of bound states
trapped in this potential well or penetrating the horizon of the Galileon black hole, was investigated. The strength of the coupling of the scalar field to Einstein tensor, which signals  how strongly matter is coupled to gravity, plays a decisive role on the behaviour of the bound states. Studying the  energies going to infinity  (the Schwarzschild limit) they form a continuum, corresponding to a continuous distribution and the absence of bound states. Moreover, the bandwidths decrease for large values of the coupling so the bound states become more and more stable while reducing the coupling renders the quasi-bound states unstable and at  some point the bound states no longer exist.

In this work we study the bound states in a wormhole geometry in the scalar-tensor Horndeski theory. These wormhole solutions  are exact static spherically symmetric solutions in the subclass of the Horndeski theory in which the scalar field is coupled kinetically to curvature. These solutions
are generated by phantom matter and the wormhole throat connects two anti-de Sitter (AdS) spacetimes \cite{Maldacena:2004rf}. Their  stability was studied in \cite{Stability}. This coupling respects the shift symmetry as in the Galileon black holes, and because of that it does not allow the scalar field to have self-interacting terms. This property leads to
the fact that the coupling constant of the scalar field to the Einstein tensor $G_{\mu\nu}$ appears directly in the metric functions of the wormhole solutions without being connected with the other parameters of the solution.

Scalar fields  around black holes and wormholes in the presence of a cosmological constant exhibit non-trivial behaviours in curved space-times. The method that gives a better understanding of the physical results and improves their accuracy is  based on the Wentzel-Kramers-Brillouin (WKB) approximation. In \cite{Grain:2006dg, Grain} stationary solutions in the Schwarzschild-Anti-de Sitter (SAdS) background were investigated and the energy levels of scalar fields were accurately obtained. In  our study using the WKB approximation  we will consider a probe scalar field of real matter scattered off in the wormhole background. We will calculate the bound states and the energy levels in this wormhole background solving semi-classically the Klein-Gordon equation.

There are various studies of field propagation in wormhole geometries. In \cite{Visser:1990wi}  an application of quantum-mechanical principles to a microscopic variant of the traversable wormholes introduced by Morris and Thorne were presented. Various aspects of transversal wormholes that they exhibit quantum teleportation  by an interaction between the two asymptotic boundaries were studied in \cite{Maldacena:2017axo}. This work was further extended in \cite{vanBreukelen:2017dul}  by studying quantum teleportation through time-shifted AdS wormholes.
In \cite{Bao:2019rjy} quantum random walks between traversable wormholes and quantum channels were studied. The effects of closed Universes that branch off or join onto an asymptotically AdS spacetime by means of an effective quantum field theory in an AdS background were studied in \cite{Barcelo:1997hm}. The problem of reversibility of wormholes in the framework of quantum  improvement  of  gravity  theory was investigated in \cite{Moti:2020whf}. Euclidean wormholes in a holographic setup were  discussed in \cite{Betzios:2019rds}.

The motivation of this work is to study the quantum mechanical effects of a wormhole in AdS spacetime in which the matter appears explicitly in the shift and lapse functions. Because  the matter which support the wormhole has a negative kinetic energy, i.e. is a phantom field,  it would be interesting to study what kind of bound states are formed in the wormhole configuration and what is their dynamics. This may help us to understand if matter is localized in the throat or if it can tunnel from one region (universe) to the second region (universe). For this reason we have chosen to work with an exact wormhole solution in the scalar-tensor Horndeski theory. However, there are some criticisms of the stability of the wormhole solutions in a asymptotically flat spacetime in the scalar-tensor Horndeski theory \cite{Evseev:2017jek}.

In our study the wormhole lives in a asymptotically AdS spacetime. Nevertheless, the stability  is an important issue in all wormhole configurations because of the presence of phantom matter. In order to study the stability of the wormhole solutions one has to calculate the gravitational  tensor (axial and polar) and the scalar perturbations of these solutions. The introduction of gravitational perturbations (such as the ones considered in \cite{Bronnikov:2004ax}-\cite{Bronnikov:2012ch}) may lead to an unstable wormhole, due to the presence of phantom matter, which can potentially expand or collapse into a black hole \cite{Gonzalez:2008xk,Doroshkevich:2008xm}. As a first step  the scalar perturbations in the background of the wormhole solution \cite{Korolev:2014hwa} was carried out in \cite{Vlachos:2021weq}. These perturbations generated echoes in  this wormhole background and it was found that they do not decay with time, but have constant and equal amplitude to that of the initial ringdown. The constancy of the amplitude of echoes is related to the absence of dissipation and may be an indication of the existence of normal oscillation modes, as well as potential instabilities. To have a better understanding of the stability issue the gravitational perturbations has to be calculated \cite{Work}.  Recently in \cite{DelAguila:2021rck}, various Morris-Thorne-like wormholes where studied and it was found that some models could be linearly stable under gravitational perturbations.

The work is organized as follows. In Section \ref{sect2} we discuss the wormhole solution presented in \cite{Korolev:2014hwa}. In Section \ref{sect3} we calculate the Regge-Wheeler potential generated in the background of this wormhole solution. In Section \ref{sect4} using the WKB approximation we calculate the bound states and the energy levels and in  Section \ref{sect5} we study the  quantum gravity effects in the extreme mass limit and we calculate the  non-resonant energies supported by the potential wells.
 In the Appendix we derive the quantization condition and in Section \ref{sect6} we conclude.

\section{Wormhole solution in the scalar-tensor Horndeski theory}
\label{sect2}

In the  Einstein theory of gravity it was shown \cite{Morris} that it is possible to find static solutions describing transversable spacetimes connecting two far away regions of our universe or even different universes. Far away from the tunnel, spacetime can be flat or described by a curved geometry. The spacetime wormhole ansatz of Morris and Thorne was formulated originally for static spherically symmetric metrics in the form
\begin{equation}
    d{s}^2 = e^{2\varPhi(r)} d{t}^2 - \frac{d{r}^2}{1-b(r)/r} -r^2 (d\theta^2  +sin^2\theta d\phi^2)~,\label{wormmetr}
\end{equation}
where $ e^{\varPhi(r)} $ and $b(r)$ are arbitrary functions of the radial coordinate.

To have a wormhole these two functions must satisfy some general constraints. These constraints give a minimum set of conditions which can lead to a geometry featuring two regions connected by a bridge:
\begin{itemize}
 \item A no-horizon condition, i.e.  $ e^{\varPhi(r)}\neq0 $, which means that $\varPhi(r)$ is finite throughout the spacetime in order to ensure the absence of horizons and singularities.

\item Minimum value of the r-coordinate, i.e. at the throat of the wormhole $r=b(r)=r_0$, $r_0$ being the minimum value of r.

\item Finiteness of the proper radial distance, i.e.
\be
\frac{b(r)}{r} \leq 1~,
\ee
for $r \geq r_0$ throughout the spacetime. The equality sign holds only at the throat.

\end{itemize}

As we already mentioned there are wormhole solutions in scalar-tensor theories. We briefly present the wormhole solution of the following gravitational theory with a  non-minimal derivative coupling (NMDC) of a scalar field to Einstein tensor presented in \cite{Korolev:2014hwa}. Consider the action
	\begin{equation}
	\label{action}
	S=\int d^4x\sqrt{-g}\left\{\frac{R}{8\pi} - \left[\varepsilon g_{\mu\nu}+ \eta
	G_{\mu\nu}\right]\varphi^{,\mu}\varphi^{,\nu}\right\}~,
	\end{equation}
	where  $\varphi$ is a real massless scalar field and $\eta$ is a parameter of nonminimal kinetic coupling with the dimension of length-squared. We note that we are using natural units such that $G=c=\hbar=1$.
	The $\varepsilon$ parameter equals $\pm1$. In the case $\varepsilon = 1$ we have a canonical scalar field with positive kinetic term, and the case $\varepsilon = -1$ describes a phantom scalar field with negative kinetic term.
	
	Following the results of \cite{Rinaldi:2012vy} the authors in \cite{Korolev:2014hwa} considered a spherically symmetric metric ansatz of the form
	\begin{equation}
	\label{metric}
	ds^2=-f(r)dt^2+g(r)dr^2+\rho^2(r)d\Omega^2~.
	\end{equation}
	Since we are  interested in spherically symmetric solutions, we can also choose $\varphi=\varphi(r)$. Under this ansatz, the equations of motion of the theory are given by
	
	\begin{subequations} \label{fieldeq1}
		\begin{align}
		\label{first}
		&\frac{\sqrt{fg}}{g}\psi\left[\varepsilon\rho^2+\eta\left(\frac{\rho\rho'f'}{fg
		}+\frac{\rho'^2}{g}-1\right)\right]=C_0~,\\
		\label{second}
		&\rho\rho'\frac{f'}{f}=\frac{
			g(g-\rho'^2)-4\pi\eta\psi^2(g-3\rho'^2)+4\pi\varepsilon \rho^2\psi^2
			g}{g-12\pi\eta\psi^2}~,\\
		\label{third}
		&\frac{\rho\rho'}{2}\left(\frac{f'}{f}-\frac{g'}{g}\right)=\frac{
			g(g-\rho'^2-\rho\rho'')+4\pi\eta\psi^2(2\rho'^2+\rho\rho'')+4\pi\eta
			\rho\rho'(\psi^2)'}{ g - 12\pi\eta\psi^2}~,
		\end{align}
	\end{subequations}
	where $C_0$ is an integration constant, and $\psi\equiv\varphi'$. Setting the integration constant $C_0=0$ \cite{Korolev:2014hwa}, one can find an exact solution to (\Ref{first}) given by
	\begin{equation}
	\label{genf}
	f(r)=\frac{C_1}{\rho}\exp\left(-\int{\frac{(\varepsilon\rho^2-\eta)g}{\eta\rho\rho'}dr}\right)~,
	\end{equation}
	where $C_1$ is an integration constant.
	Using this result, along with (\ref{second}), one can also derive $\psi^2$
	\begin{equation}
	\label{genpsi}
	\psi^2(r)=\frac{\varepsilon \rho^2 g}{8\pi\eta(\varepsilon\rho^2-\eta)}~.
	\end{equation}
	
	Making use of equations (\Ref{genf}) and (\Ref{genpsi}), the solutions to the equations of motion, (\Ref{fieldeq1}), are given by the following branches:
	
	{\bf A. $\varepsilon\eta>0$.}
	\begin{align}\label{gA}
	f(r)&=\frac{C_1}{\rho}\exp\left(-\int{\frac{(\varepsilon\rho^2-\eta)g}{\eta\rho\rho'}dr}\right)~,\\
	g(r) &= \frac{\rho'^2(\rho^2-2l_{\eta}^2)^2}{(\rho^2-l_{\eta}^2)^2 F(r)}~,\\
	\label{FA}
	F(r) &=
	3-\frac{8m}{\rho}-\frac{\rho^2}{3l_\eta^2}+\frac{l_\eta}{\rho}\arctanh\frac{\rho}{l_\eta}~.
	\end{align}

	{\bf B. $\varepsilon\eta<0$.}
	\begin{align}\label{gB}
	f(r)&=\frac{C_1}{\rho}\exp\left(-\int{\frac{(\varepsilon\rho^2-\eta)g}{\eta\rho\rho'}dr}\right)~,\\
	g(r) &=\frac{\rho'^2(\rho^2+2l_{\eta}^2)^2}{(\rho^2+l_{\eta}^2)^2 F(r)}~,\\
	\label{FB}
	F(r) &=
	3-\frac{8m}{\rho}+\frac{\rho^2}{3l_\eta^2}+\frac{l_\eta}{\rho}\arctan\frac{\rho}{l_\eta}~.
	\end{align}
	Here $m$ is an integration constant and $l_\eta=|\varepsilon\eta|^{1/2}$ is a characteristic scale of the nonminimal kinetic coupling.
	
	The above equations suggest that we are free to choose the coefficient of the $S^2$ hypersurface in the metric.   Korolev and Sushkov \cite{Korolev:2014hwa} proposed a wormhole configuration of the above solution of the form:
	\begin{equation}\label{throat}
	\rho(r)=\sqrt{r^2+a^2}~,
	\end{equation}
	where $a>0$ is a free parameter. If $f(r)$ and $g(r)$ are everywhere positive and regular function with a domain $r\in(-\infty,\infty)$, then the solution describes a wormhole with a throat at $r=0$, while the parameter $a$ is just the throat radius. The throat at $r=0$ could be misleading. It is well known that the corresponding $S^2$ hypersurfaces of the embedded spacetime should be decreasing as one approaches the wormhole. \Ref{throat} has been chosen such that there is always a minimum $S^2$ hypersurface with non-zero radius. As such, the throat at $r=0$ can be straightforwardly understood from the chosen wormhole configuration, \Ref{throat}. An important note here is the fact that the coordinates $(t,r,\theta,\phi)$ are not the Schwarzschild coordinates since $r$ is not the curvature radius of a coordinate sphere $r=Const>0$.
	
	Substituting $\rho(r)=\sqrt{r^2+a^2}$ into the formulas (\Ref{genpsi}), (\Ref{gA})-(\Ref{FB}), we derive the solutions for $g(r)$ and $\psi^2(r)$ in an explicit form. The solution (\Ref{genf}) for $f(r)$ contains the indefinite integral, which in this case cannot be expressed in terms of elementary functions. From the two branches of solutions, only the case where $\varepsilon\eta<0$ can be considered physical \cite{Korolev:2014hwa}. In particular, one needs to set $\epsilon=-1$ and $\eta>0$ \cite{Korolev:2014hwa}. In this case,  by substituting $\rho(r)=\sqrt{r^2+a^2}$ into the formulas (\Ref{gB})-(\Ref{FB}) and (\Ref{genf}), we obtain the following solutions:
	\begin{align}
	g(r) &= \frac{r^2(r^2+a^2+2l_\eta^2)^2}{(r^2+a^2)(r^2+a^2+l_\eta^2)^2 F(r)}, \label{gwh}~,\\
	f(r) &= \frac{a}{\sqrt{r^2+a^2}}\exp\left[\int_0^r\frac{r(r^2+a^2+2l_\eta^2)^2} {l_\eta^2(r^2+a^2)(r^2+a^2+l_\eta^2)F(r)}dr\right]~, \label{fwh}\\
	\psi^2(r) &= -\frac{\varepsilon}{8\pi l_\eta^2}
	\frac{r^2(r^2+a^2+2l_\eta^2)^2}{(r^2+a^2)(r^2+a^2+l_\eta^2)^3 F(r)}~, \label{psiwh}
	\end{align}
	where
	\begin{equation}
	F(r)=3-\frac{8m}{\sqrt{r^2+a^2}}+\frac{r^2+a^2}{3l_\eta^2}
	+\frac{l_\eta}{\sqrt{r^2+a^2}}\arctan \left(\frac{\sqrt{r^2+a^2}}{l_\eta}\right)~,
	\end{equation}
	and the integration constant $C_1=a$ in the expression for $f(r)$ is chosen so that $f(0)=1$. The function $F(r)$ has a minimum at $r=0$, thus, to make it everywhere positive, it is sufficient to demand $F(0)>0$. Hence one can derive the limitation on the upper value of the parameter $m$
	\begin{equation}\label{M}
	m<\frac{a}{2}\left(\frac34+\frac{\alpha^2}{12}+\frac{1}{4\alpha}\arctan\alpha\right)=M_{crit}~,
	\end{equation}
	where $\alpha\equiv a/l_\eta$ is the dimensionless parameter which defines the ratio of two characteristic sizes: the wormhole throat radius $a$ and the scale of nonminimal kinetic coupling $l_\eta$. In the particular case $a\ll l_\eta$ we get $2m<a$. Furthermore, we assume that the value of $m$ satisfies the condition (\Ref{M}), and therefore the function $F(r)$ is positive definite, i.e. $F(r)>0$. An important note is that the solutions (\Ref{gwh})-(\Ref{fwh}) of the theory correspond to two AdS-asymptotic spacetimes.

In GR to generate a wormhole solution we choose a static spherically symmetric metric, and as we already discussed,  conditions on this metric function  were imposed. However, the condition on the wormhole throat leads to the violation of NEC which implies the presence of exotic matter. The amount of exotic matter which is required for the formation of the wormhole depends on the spacetime geometry  and in \cite{Visser:2003yf}   wormhole geometries were found which  are supported by arbitrarily small quantities of exotic matter. In the scalar-tensor theory we consider, the information of the presence of exotic matter appears explicitly in the metric function (see rel. (\ref{gB})) and the amount of exotic matter present in the wormhole geometry depends on how strong is the  coupling of matter to curvature through the coupling $\eta$ of the scalar field to the Einstein tensor.

	\section{The Regge-Wheeler Potential}
\label{sect3}
	
	Having found the solutions of the field equations, we may continue with the main problem at hand. We wish to study the propagation of a test scalar field in the vicinity of the wormhole. In particular, our goal is to derive the energy eigenstates of the scalar field in the wormhole and study the transmission amplitudes between the two regions. In order for these bound states to exist, the Klein-Gordon equation in the wormhole background has to exhibit a radial potential containing at least one local well.
	
	The Klein-Gordon equation of motion for a test massless scalar field $\Phi$ in a spherically symmetric curved background reads
	\begin{equation}
		\label{KG}
		\frac{1}{\sqrt{-g}}\partial_\mu\left[
		\sqrt{-g}g^{\mu\nu}\partial_\nu\Phi\right]=0~.
	\end{equation}
	We choose the ansatz of $\Phi(t,r,\theta,\phi)=R(r)
	Y^l_m(\theta,\phi)e^{-iwt}$ to disentangle the radial, angular and temporal parts of the field. Using the above ansatz we can reduce the differential equation of motion to the following ODE
	\begin{equation}
	\label{KG1}
	h(r)\partial_{r}\left[\rho^2(r)h(r)\partial_{r}R(r)\right]+[w^2-l(l+1)f(r)]R(r)=0~,
	\end{equation}
	where we set $h(r)=\frac{\sqrt{f(r)}}{\sqrt{g(r)}}$.
	Introducing the tortoise coordinate $r^*$,
	\begin{equation}
	\label{tort}
	dr^*=\frac{dr}{h(r)}\implies h(r)\frac{d}{dr}=\frac{d}{dr^*}~,
	\end{equation}
	we simplify the above equation to
	\begin{equation}
	\label{KG2}
	\partial_{r^*}\left[\rho^2(r)\partial_{r^*}R(r)\right]+[w^2-l(l+1)f(r)]R(r)=0~.
	\end{equation}
	The next step is to perform the substitution $R(r)=\frac{u(r)}{\rho(r)}$. The radial equation, after some simple algebra, takes the following form,
	\begin{equation}
		\label{KG3}
		\frac{\partial^2 u(r)}{\partial r^{*2}}+\left[w^2-V^2_{RW}(r)\right]u(r)=0~,
	\end{equation}
	where $V^2_{RW}$ is interpreted as the squared Regge-Wheeler potential, so as to recover the Hamilton-Jacobi equation. The explicit form of the potential reads
	\begin{equation}
	\label{potential}
	V^2_{RW}=l(l+1)\frac{f(r)}{\rho^2(r)}+\frac{2f(r)g(r)\frac{\partial^2\rho(r)}{\partial r^2}+g(r)\frac{\partial f(r)}{\partial r}\frac{\partial \rho(r)}{\partial r}-f(r)\frac{\partial g(r)}{\partial r}\frac{\partial \rho(r)}{\partial r}}{2g(r)\rho^2(r)}~.
	\end{equation}
	Let us note that equation (\Ref{potential}) is a general solution for the Regge-Wheeler potential of a spherically symmetric metric.	In general, we shall be looking for stationary state solutions of the Klein-Gordon equation. Equation (\Ref{KG3}) will be solved using the WKB approximation which we review in the following section.
	
	\section{WKB approximation}
\label{sect4}

	The WKB approximation is a powerful semi-analytical method for solving linear differential equations whose highest derivative is multiplied by a small parameter $\epsilon$. Let us consider the following second order differential equation
	\begin{equation}
	\label{wkb}
		\epsilon^2 y(x)''=Q(x)y(x)~.	\end{equation}
	The WKB formula consists of an asymptotic approximation to $y(x)$ of the following form
	\begin{equation}
		\label{app}
		y(x)\sim\exp\left[\frac{1}{\delta}\sum_{n=0}^{\infty}\frac{S_n(x)}{\delta^n}\right]~.
	\end{equation}
	This expression is the starting formula from which all WKB approximations are derived.
	Differentiating (\Ref{app}) twice yields
	\begin{equation}
	\label{dif}
		y''\sim\left[\frac{1}{\delta^2}\left(\sum_{n=0}^{\infty}\delta^n S'_n\right)^2+\frac{1}{\delta}\sum_{n=0}^{\infty}\delta^nS''_n\right]y~.
	\end{equation}
	Plugging (\Ref{dif}) into (\Ref{wkb}), one finds that
	\begin{equation}
		\label{seq1}
		\frac{\epsilon^2}{\delta^2}S_{0}^{'2}+\frac{2\epsilon^2}{\delta}S'_0S'_1+\frac{\epsilon^2}{\delta}S''_0+...=Q(x)~,
	\end{equation}
	which provides the solution for each $S_n$ by equating terms of the same order. Setting $\delta=\epsilon$, one finds the following recursive sequence of equations
	\begin{align}
	\label{seq}
		&S_0^{'2}=Q(x)~,\\
		&2S'_0S'_1+S_0''=0~,\\
		&2S'_0S'_n+S_{n-1}''+\sum_{j=1}^{n-1}S'_jS'_{n-j}=0, \qquad n\geq2~.
	\end{align}
	Up to the second order, making use of the first two equations in (\Ref{seq}), one finds the eikonal and transport equations
	\begin{align}
		&S_0(x)=\pm\int^{x}\sqrt{Q(x')}dx'~,\\
		&S_1(x)=-\frac{1}{4}\ln{Q(x)}~.
	\end{align}
	The formal WKB approximation of equation (\Ref{wkb}) is the linear combination of the solutions and reads, to first order:
	\begin{equation}
	\label{sol}
		y(x)\sim A Q^{-1/4}(x)\exp\left[\frac{1}{\epsilon}\int^{x}\sqrt{Q(x')}dx'\right]+BQ^{-1/4}(x)\exp\left[-\frac{1}{\epsilon}\int^{x}\sqrt{Q(x')}dx'\right]~.
	\end{equation}	
	The factors $A$ and $B$ are integration constants.

	WKB has been extensively used in the study of dissipative and dispersive phenomena, in finding the energy eigenvalues of a wavefunction in a potential well, in extracting the quasinormal frequencies of black holes, etc. A deep and detailed analysis of this method can be found in \cite{bender}. An application of WKB in quantum mechanical phenomena can be found in \cite{merzbacher}, \cite{zettili}. We will use the WKB approximation to find quantum bound states trapped in the vicinity of the AdS-wormhole potential barriers. In the figure below, we show the Regge-Wheeler potential of the NMDC Wormhole for different values of the orbital quantum number.
		
\begin{center}
		\begin{figure}[h!]
			\includegraphics[width=100mm,scale=0.5]{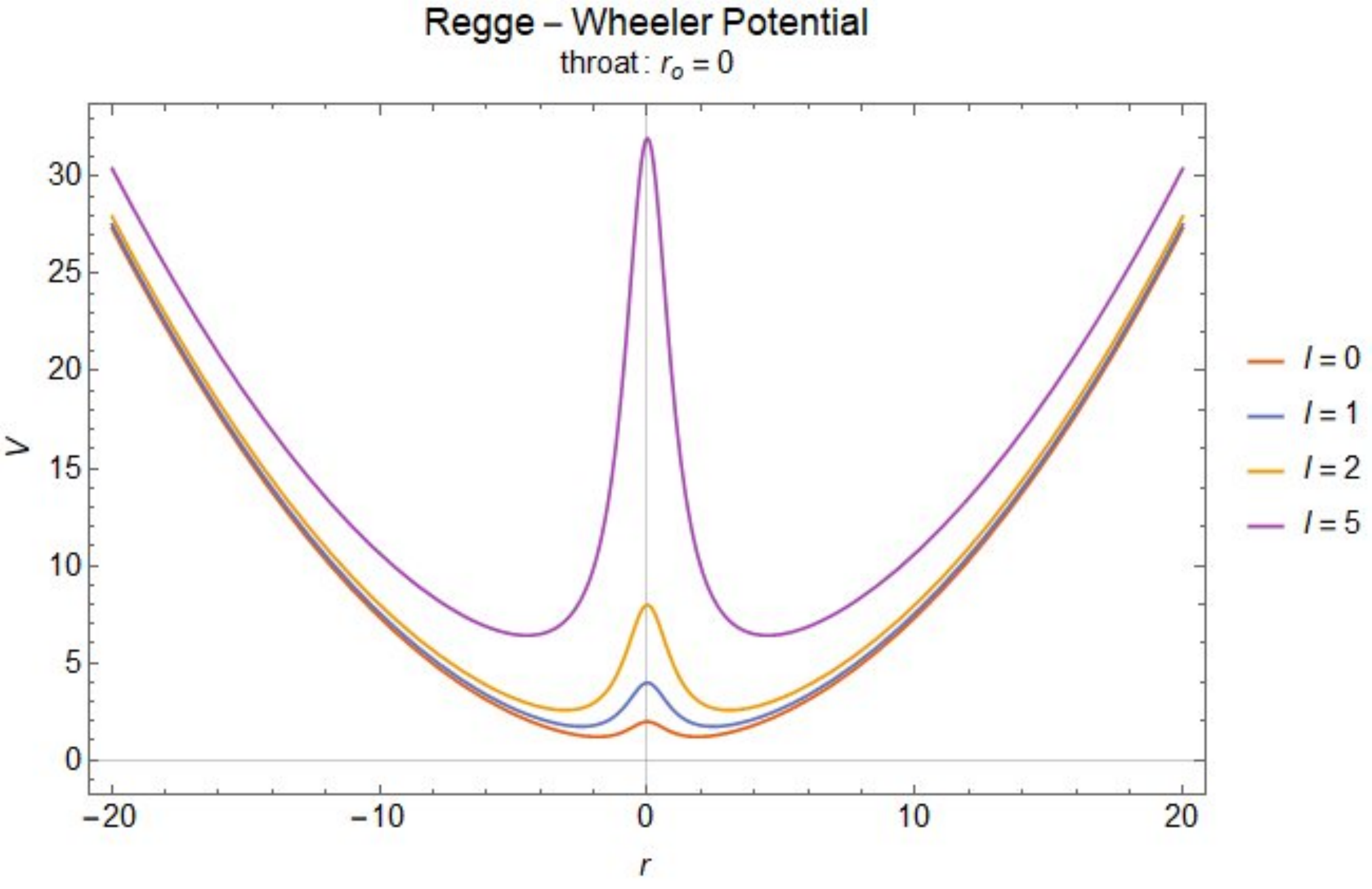}
			\caption{Regge-Wheeler Potential for different values of the orbital quantum number. The throat is located at $r=0$. We have fixed the other parameters to $m=0.1,a=1,l_\eta=1$}. \label{fig1}
		\end{figure}
	\end{center}

		As can be seen from Fig. \ref{fig1}, high values of the orbital quantum number result in the creation of two distinct potential wells, one for each region. Thus a scalar field trapped in one region can tunnel to the second region through the potential peak on the throat. The final result is of course the formation of bound states and the problem parallels that of a particle wave function in the potential created by the nuclei in a diatomic molecule. It is clear from the form of the potential that the transmission amplitude of the tunnelling effect is highly dependent on the orbital number of the test scalar field. As such, since for low orbital numbers the potential peak at the throat is small, the wormhole allows free passage of the  states with low orbital quantum numbers. Therefore, we concludes that the throat acts as a low-pass filter on the transmission of the test wave. However, due to the tunnelling effects we are considering, this is not exactly the case. The flow between the two regions is indeed inhibited by the contribution of the states with high orbital quantum numbers, but if one waits long enough, the current density is seen to oscillate between confinement in each region.
	
\begin{center}
		\begin{figure}[h!]
			\includegraphics[width=100mm,scale=0.5]{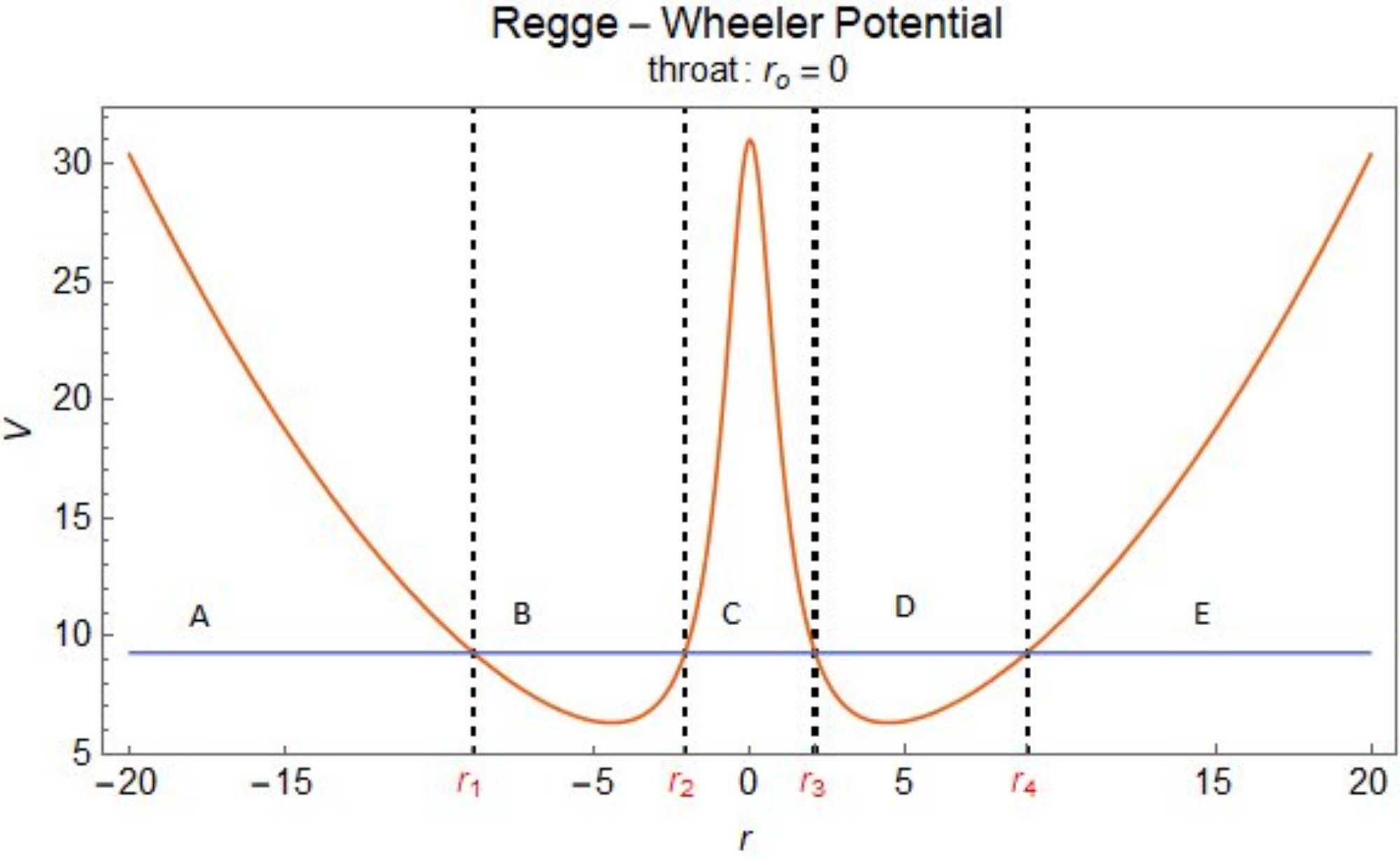}
			\caption{Regions for the calculation of wave functions.} \label{fig2}
		\end{figure}
	\end{center}

	 In the following, we will be referring to the potential peak of the throat as $V_{max}$. Let us consider a frequency value of the scalar field, $w^2\in(V_{min},V_{max})$, that cuts the potential four times, as depicted in Fig. \ref{fig2}. 	
	 	
	 The four turning points are denoted as $r_i$ for $i=1,2,3,4$. This results in five regions, labelled $A$ to $E$ from left to right, where the  WKB solution changes branch. The amplitudes of the WKB wave functions can be derived from continuity boundary conditions on the turning points. Since, however, the WKB approximation diverges there, one needs to perform a Taylor expansion of the potential in the vicinity of the turning points and solve the corresponding differential equation, which results in the Airy functions. Then, one needs to take the asymptotic limit of the Airy function solutions and match them with the WKB wave functions. The matching procedure can be found in the appendix of \cite{Grain:2006dg}. It is clear that for frequencies above the maximum of the potential at the throat, the problem is reduced to finding energy eigenstates in a potential well.
	
	 Depending on the orbital number of the scalar field, there is a different number of energy eigenvalues which lie below the potential maximum at the throat. In particular, fixing the values of the parameters to $m=0.1,a=1,l_{\eta}=1$, one finds that s-waves can traverse the wormhole freely and that there are no bound states below $V_{max}$. In the following, we present the solutions for the test scalar field, whose energy is small enough in order to interact with the throat and we have fixed the parameters to $m=0.1,a=1,l_{\eta}=1$.
	
	  The WKB solutions for each region read	\begin{subequations}
		\begin{align}
		\label{solutions}
		&U_A(r^*)=\frac{N(-1)^n}{\sqrt{|p(r^*)|}}\exp\left[-\int^{r^*_1}_{r^*}|p(r^{*'})|dr^{*'}\right]~,\\
		&U_B(r^*)=\frac{2N(-1)^n}{\sqrt{p(r^*)}}\sin\left(\int^{r^*}_{r^*_1}p(r^{*'})dr^{*'}+\pi/4\right)~,\\
		&U_C(r^*)=\frac{2N \cos\xi}{\sqrt{|p(r^*)|}}\exp\left[\int^{r^*_3}_{r^*}|p(r^{*'})|dr^{*'}\right]+\frac{N\sin\xi}{\sqrt{|p(r^*)|}}\exp\left[-\int^{r^*_3}_{r^*}|p(r^{*'})|dr^{*'}\right]~,\\
		&U_D(r^*)=\frac{2N}{\sqrt{p(r^*)}}\sin\left(\int^{r^*_4}_{r^*}p(r^{*'})dr^{*'}+\pi/4\right)~,\\
		&U_E(r^*)=\frac{N}{\sqrt{|p(r^*)|}}\exp\left[-\int^{r^*}_{r^*_4}|p(r^{*'})|dr^{*'}\right]~,
		\end{align}
	\end{subequations}
	with the quantization condition
	\begin{equation}
	\label{quan}
	\tan{\xi}=\pm 2e^{\zeta}~,
	\end{equation}
	where the $\pm$ factor corresponds to even$(+)$ and odd$(-)$ energy level numbers, $n$  and as we will show later its connection to  the well known Bohr-Sommerfeld quantization condition \cite{zettili}.
	The various factors found in the solutions are defined as follows
	\begin{eqnarray}
	\label{defs}
	p(r^*)&\equiv&\sqrt{w^2-V^2_{RW}(r^*)}~,\\
	|p(r^*)|&\equiv&\sqrt{V^2_{RW}(r^*)-w^2}~,\\
	\xi&\equiv&\int_{r_3^*}^{r_4^*} p(r^{*'})dr^{*'}~,\\
	\z&\equiv&\int_{r^*_2}^{r^*_3}|p(r^{*'})|dr^{*'}~.
	\end{eqnarray}
	The quantization condition (\Ref{quan}) can be derived in a straightforward fashion from the scattering matrix of the fields at the throat. The relations between the solutions in regions B and D, which define the scattering matrix, are worked out in the Appendix. The result reads
	 \begin{equation}
	 \label{smat}
	\begin{pmatrix}
	U_{B_{in}} \\ U_{B_{out}}
	\end{pmatrix}
	=
	\begin{pmatrix}
	e^{\zeta} + \frac{1}{4}e^{-\zeta} & i(e^{\zeta} - \frac{1}{4}e^{-\zeta})\\
	-i (e^{\zeta} - \frac{1}{4}e^{-\zeta}) & e^{\zeta} + \frac{1}{4}e^{-\zeta}
	\end{pmatrix}
	\begin{pmatrix}
	U_{D_{out}} \\ U_{D_{in}}
	\end{pmatrix}\\
	=S\begin{pmatrix}
	U_{D_{out}} \\ U_{D_{in}}
	\end{pmatrix}~,\\
	\end{equation}
		where $U_{B_{in}}$ is the incoming wave from the left region to the throat, while $U_{B_{out}}$ is the outgoing wave to the left region. Similarly, $U_{D_{out}}$ is the outgoing wave to the right region, while $U_{D_{in}}$ is the incoming wave from the right region to the throat. The functions $U_{B_{in}},\ U_{B_{out}},\ U_{D_{out}}$ and $U_{D_{in}}$ are the WKB solutions with the lower limit of the integral fixed on the throat turning point. In summary: $U_{B_{in}}$ and $U_{D_{out}}$ move to the right, while $U_{B_{out}}$ and $U_{D_{in}}$ move to the left. The transmission amplitude is given by
		\begin{equation}\label{Trans}
		T=\frac{1}{|S_{11}|^2}\approx e^{-2\zeta}~.
		\end{equation} Starting with the scattering matrix one may also prove, as explained in the appendix, that
	\begin{equation}\label{smat2}
		S_{11}e^{-2i\xi}i+S_{12}=-iS_{22}e^{2i\xi}+S_{21}~,
	\end{equation}
	which leads to (\Ref{quan}). We observe that the solutions are phase shifted at the AdS-barrier turning point by $e^{i\xi},$ which gives rise to the $e^{2i\xi}$ terms. On the other hand the reflection on the AdS barrier, i.e. the asymptotic behaviour of the potential, imposes a $\pi/2$ phase difference between the incoming and the outgoing wave: this lies at the origin of the factors of $i.$

\begin{center}
	 	\begin{figure}[h!]
	 		\includegraphics[width=80mm,scale=0.1]{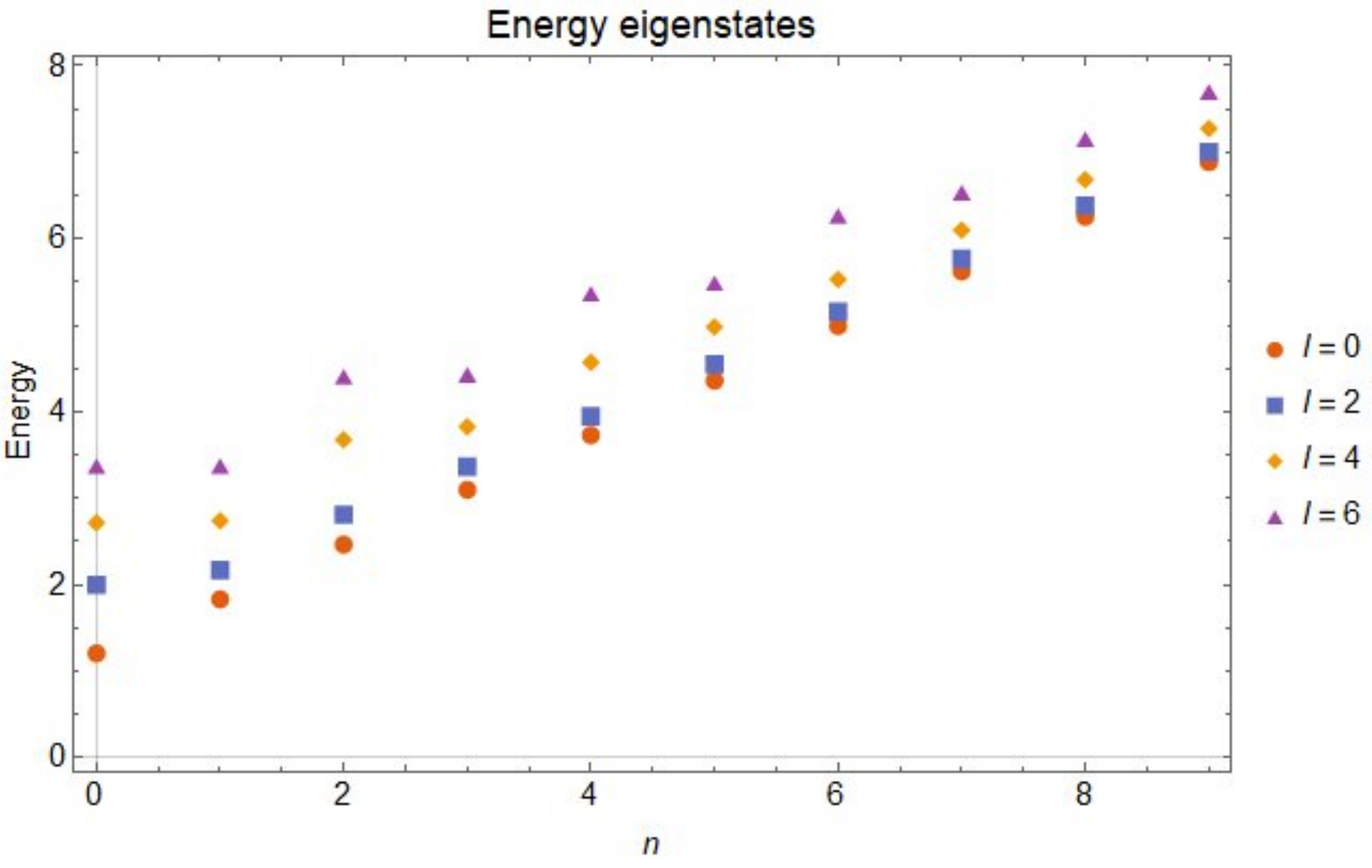}
	 		\includegraphics[width=80mm,scale=0.12]{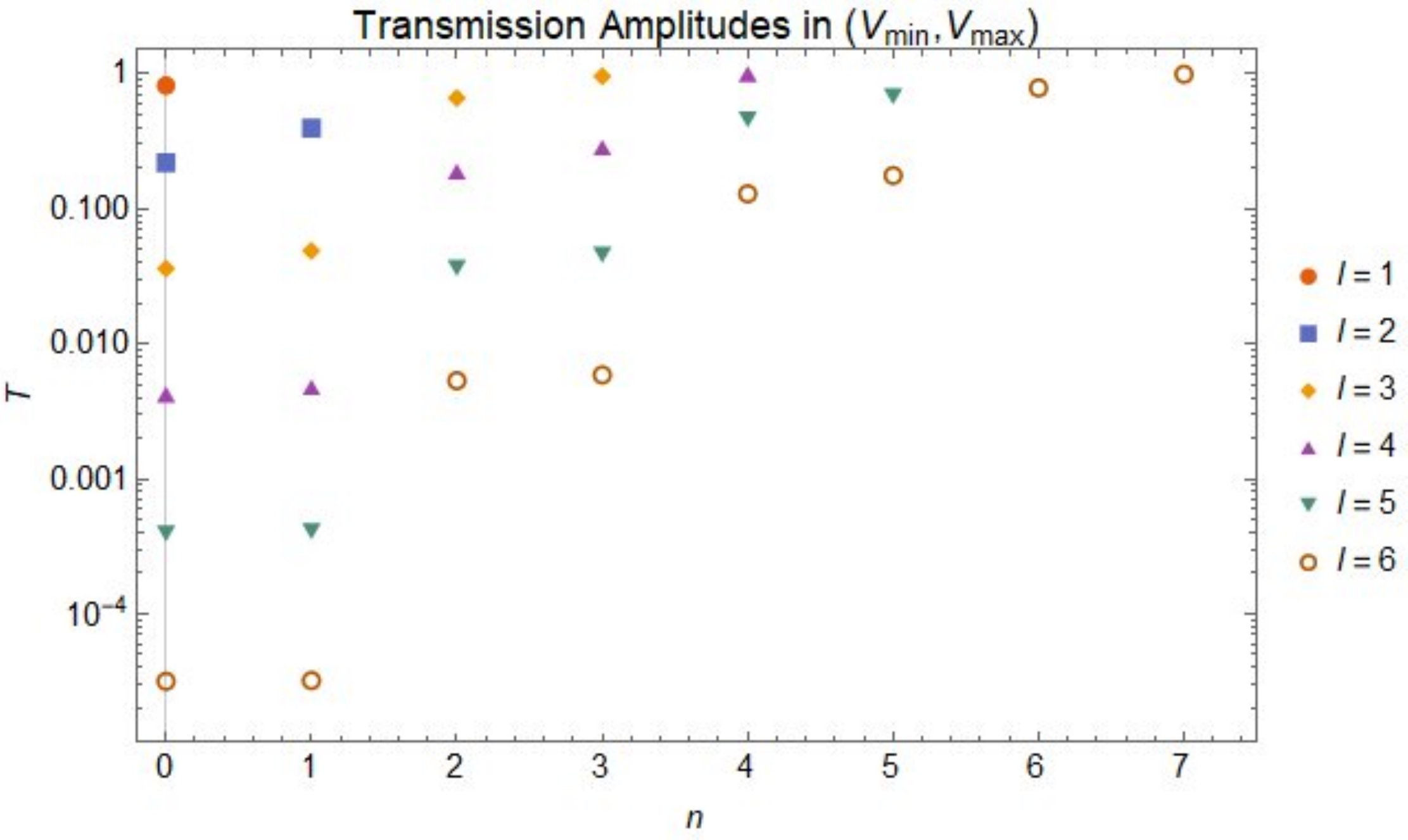}
	 		\caption{(a) Ten lowest energy levels in the Regge-Wheeler potential for the values $l=0,\ 2,\ 4,\ 6$ of the orbital quantum number $l.$ For $l=6$ one gets the highest energy levels for each value of $n,$ while the lowest ones appear for $l=0.$ (b) Transmission amplitudes for the energy eigenstates to pass the throat for various values of the orbital quantum number. The orbital quantum number $l=1$ has the maximum transmission amplitudes for each $n,$ while $l=6$ corresponds to the minimal ones. For $l=0$ there is no barrier, so that transmission amplitude does not make sense in this case.} \label{entrans}
	 	\end{figure}
	 \end{center}

	As we already discussed, Fig. \ref{fig1} shows that for high orbital numbers for the scalar field, the Regge-Wheeler potential develops a potential peak around the throat and two potential wells are formed, while for low orbital numbers the potential peak at the throat is small and the potential wells are flattened. Therefore, we expect that the behaviour of the scalar field passing through the throat, to depend on the orbital numbers of the scalar wave. In Fig. \ref{entrans} we depict the energy levels and the transmission amplitudes for the energy eigenstates of the scalar field passing the throat.

	In particular in  the left panel of Fig. \ref{entrans}  the results for the energy eigenstates of the scalar field are shown.  We see that the lowest lying energy levels are almost  degenerate for high values of the orbital number. For example if we concentrate on the data on the top, corresponding to $l=6,$ it is easily seen that the energy level for $n=0$ and $n=1$ are approximately the same, and similar conclusions hold for the couple $(n=2,\ n=3).$ For larger values of $n$, this quasi-degeneracy is less pronounced. This effect does not hold for lower orbital states: for example no sign of degeneracy is present for $l=0,$ which corresponds to the data set lying lowest in the figure. This provides some indication that the throat indeed acts as a (quasi) low-pass filter.
	
	In the right panel of Fig. \ref{entrans} we present a logarithmic plot of the transmission amplitudes associated with energy values calculated previously. The points in Fig. \ref{entrans} are different transmission amplitudes calculated from (\Ref{Trans}), when one plugs in the WKB energy levels found previously. As one can see from the figure, the transmission amplitudes for high orbital states are several orders of magnitude smaller than the ones for the low orbital quantum numbers. This is another indication that the throat blocks the high orbital states from moving between the two regions.
	
	To summarize our findings shown in Fig. \ref{entrans}:  the energy levels with fixed $n$ increase monotonically with $l,$ and  as the energy level $n$ grows larger, the eigenstates of the scalar field do not depend very much on the orbital number $l$ and tend to a common value.

 The question is how we can understand the quasi-degeneracy of the  orbital energy levels we observed in  Fig. \ref{entrans}. To see this, let us express the quantization condition, (\Ref{quan}), as $\cot\xi=\pm\frac{1}{2}e^{-\zeta}.$ When the orbital state of the test scalar field is high, the potential peak on the throat is very wide, which results in high values for $e^{\zeta}$.  In this regime, $e^{-\zeta}$ is of course small, and the quantization condition can be rewritten as
 \begin{equation}
 \label{qquan}
 \cot\xi\approx-(\xi-(n+\frac{1}{2})\pi)=\pm\frac{1}{2}e^{-\zeta}\rightarrow \xi\approx (n+\frac{1}{2})\pi\mp \frac{1}{2}e^{-\zeta}~.
 \end{equation}
 The first part of the equation, $\xi\approx (n+\frac{1}{2})\pi$, is the well known Bohr-Sommerfeld quantization condition for a particle trapped in a potential well. The second well that is introduced due to the wormhole geometry, splits the energy levels by a factor analogous to the potential barrier on the throat. This means that the energy levels shown in figure \ref{entrans} can be approximated by solving for $w^2$ the condition $\xi\approx \frac{1}{2}\pi\mp \frac{1}{2}e^{-\zeta}$ for the first couple of energy levels, $\xi\approx \frac{3}{2}\pi\mp \frac{1}{2}e^{-\zeta}$ for the second couple of energy levels, etc, while the Bohr-Sommerfeld quantization condition would result in degenerate states.  These corrections are not important for too large values of $l.$	Therefore, the appearance of the low energy levels in close pairs in Fig. \ref{entrans} (left panel) is due to the quantization condition, (\Ref{quan}) and  in a sense  the wormhole geometry imposes corrections to the energy levels of the scalar field.

 \begin{center}
			\begin{figure}[h!]				
			    \includegraphics[width=70mm,scale=0.5]{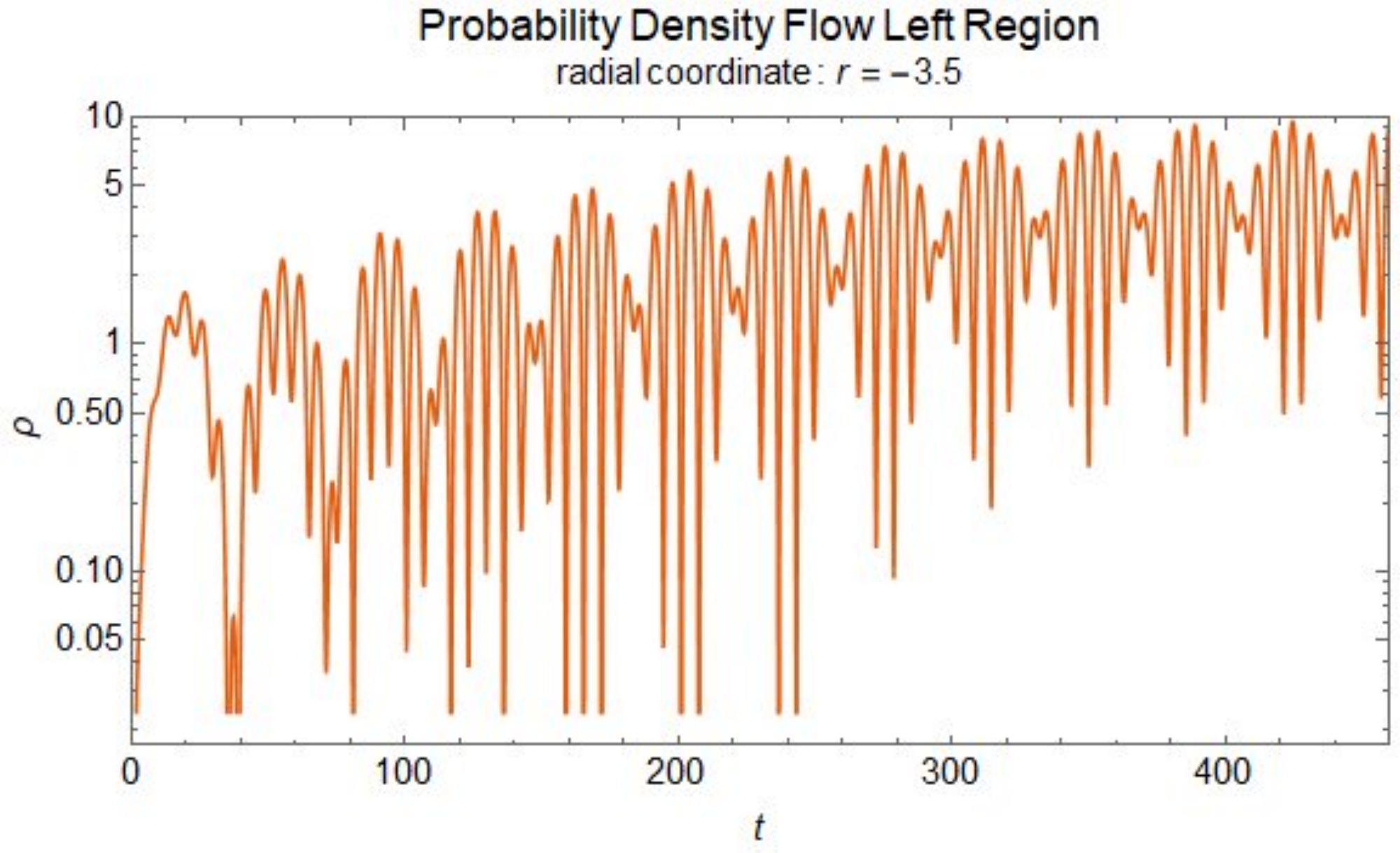}	
				\includegraphics[width=70mm,scale=0.5]{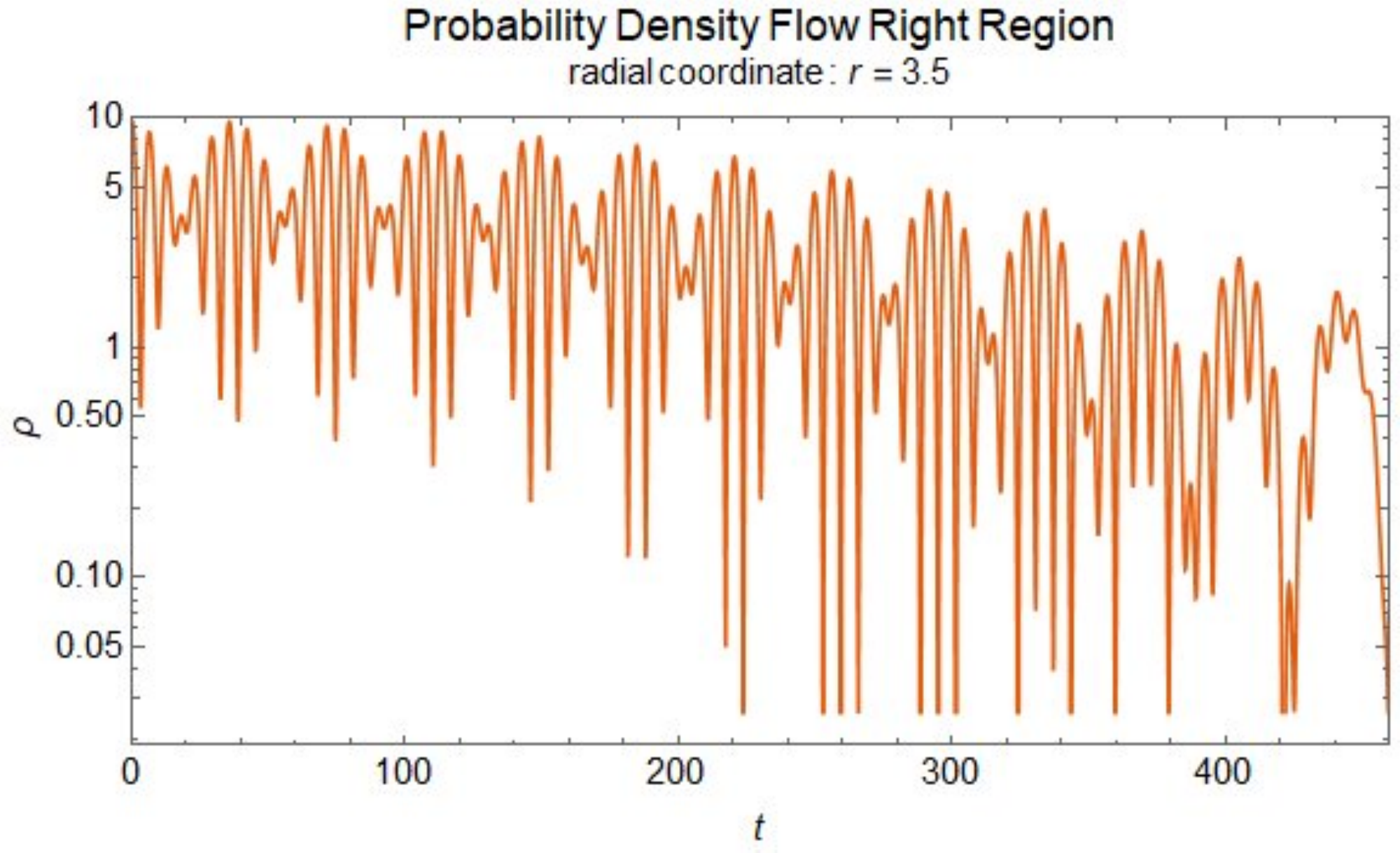}
				\caption{Flow in the left  region (left panel) for a scalar field in a superposition of $l=2$ and $l=5.$ The time range is equal to half a period. The right panel depicts the corresponding behaviour for the right hand region.}\label{RU25}
			\end{figure}
		\end{center}

 	Our results so far can be better understood by studying the zero component of the current associated with the scalar field. We recall the definition of the relevant component:
 	\be \rho=i(\Phi^*\partial_t\Phi-\Phi\partial_t\Phi^*)~,\label{dens}\ee
 	to detect  whether the scalar field feels the presence of the throat of the wormhole or not. As it is well known, the quantity (\ref{dens}) is not positive definite, which is due to the negative energy solutions that are contained in the Klein-Gordon equation. However, due to the positive definite nature of the effective potential, we keep only the positive frequency solutions, thus restricting the range of (\ref{dens}) in $\mathbb{R}^+$.
 	
We consider a superposition of various energy eigenstates with a common value for $l$
	 \begin{equation}
	 \label{ansatz}
	 \Phi(t,r,\theta,\phi)=\sum_{n,m} c_{n} R_{nl}(r)Y_{m}^l(\theta,\phi)e^{-i w_{nl}t}~.
	 \end{equation}
Now, let us consider the superposition of the fundamental and the first excited energy level for some fixed orbital number $l.$  This would result in a scalar field solution,
	 \begin{equation}
	 \label{field}
	 \Phi\sim c_1 R_{n_1 l}(r) Y_{m}^l(\theta,\phi)e^{-i w_{n_1 l}t}+c_2 R_{n_2 l}(r) Y_{m}^l(\theta,\phi)e^{-i w_{n_2 l}t}~.
	 \end{equation}
	 For definiteness we restrict our attention to the north pole of the field, $\theta=0$, where we recall that $Y_{m}^l(0,\phi)=\sqrt{\frac{2l+1}{4\pi}}\delta^0_m.$  As such, the solution takes the following general form
	 \begin{equation}
	 	\label{f}
	 	\Phi \sim c_1 R_{n_1 l}(r_{min}) Y_{m}^l(\theta=0,\phi)e^{-i w_{n_1 l}t}+c_2 R_{n_2 l}(r_{min}) Y_{m}^l(\theta=0,\phi)e^{-i w_{n_2 l}t} \sim K_1(l) e^{-i w_{n_1l}t}+K_2(l) e^{-i w_{n_2l}t}~,
	 \end{equation}
where $$K_{i}(l)=c_i R_{n_i l}(r_{min}) Y_{m}^l(\theta=0,\phi)\equiv a_i e^{i \delta_i}~.$$
Then, equation (\Ref{dens}) reads after some  algebra
	 \begin{equation}
	 \label{prob}
	 \rho=2w_{n_1 l} a_1^2 +2w_{n_2 l}a_2^2+2(w_{n_1l}+w_{n_2l}) a_1 a_2 \cos[(w_{n_2l}-w_{n_1l})t-\delta],\ \delta\equiv \delta_1-\delta_2~.
	 \end{equation}
   An important point here is that, since WKB breaks down on the turning points, (\Ref{prob}) is valid only in regions away from them. Taking that into account  we depict the graphs at the fixed point, $r_{min},$ where the potential is close to a minimum. This is because each orbital state propagates in a potential with different minima points.\\
Equation (\ref{prob}) shows that the time evolution of the flow depends explicitly on the difference of the two energy levels under consideration. Hence it is clear that, for the quasi degenerate states, i.e. the lowest energy eigenvalues of the field at large $l,$ the frequency of the flow is a very small number, resulting in the particle living in one region for large periods of time. For states, for which $w_{n_2l}-w_{n_1l}$ is relatively large (for example, for small $l$), the motion between the two regions is rapid.

The strategy we chose was to fine tune the constants $K_1$ and $K_2$ in such a way that, at $t=0,$ in the region on the right is assigned a non-zero value for $\rho,$ while the region on the left has as small a value for $\rho$ as possible. This has been done to simulate a particle starting its motion from the right ((universe)) and move towards the left one.

The time evolution for a superposition of two states with a common $l$ behaves exactly as predicted by equation (\ref{prob}), so there is no point reproducing it here. The interesting situation comes when we consider localised states, which are constructed from superposing states with, say, two different values of $l.$ In this case the states of small $l,$ if they are alone, move rapidly between the two regions, while the states with large $l$ move slowly. The question arises what happens when both of them are present. The behaviour is complicated by the fact that, apart from the  differences $w_{n_2l_1}-w_{n_1l_1}$ and $w_{n_4l_2}-w_{n_3l_2},$ characterizing the two values $l_1$ and $l_2,$ there exist contributions with the frequencies $w_{n_4l_2}-w_{n_1l_1},\ w_{n_3l_2}-w_{n_1l_1},\ w_{n_4l_2}-w_{n_3l_1}$ and $w_{n_4l_2}-w_{n_2l_1}.$

We choose to construct localised states as a superposition of $l=2$ and $l=5$ states (each one containing a superposition of the fundamental and the first excited eigenstate) to get an almost zero in the left region at $t=0.$ In figure \ref{RU25} we depict a logarithmic plot of the flow. The region on the left is empty at $t=0,$ while its density increases, in a complicated fashion, as time passes. On the contrary, the right hand region is depleted for later times. In addition, the period of the superposition is equal to the period associated with $l=5,$ which imposes a slow motion. The contribution of $l=2$ generates the very rapid oscillations within the dominant slow evolution.

\begin{center}
			\begin{figure}[h!]
			    \includegraphics[width=70mm,scale=0.5]{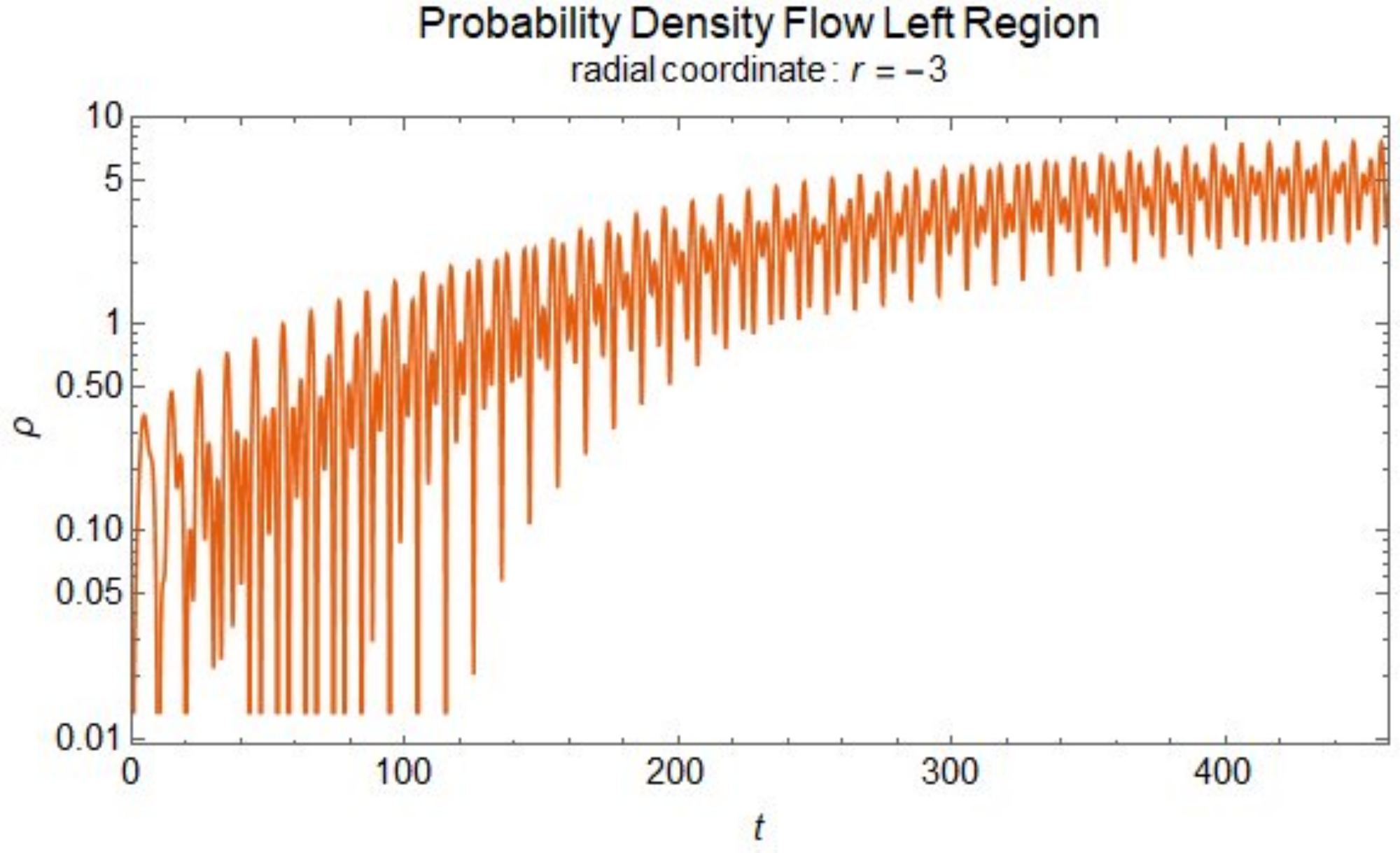}
				\includegraphics[width=70mm,scale=0.5]{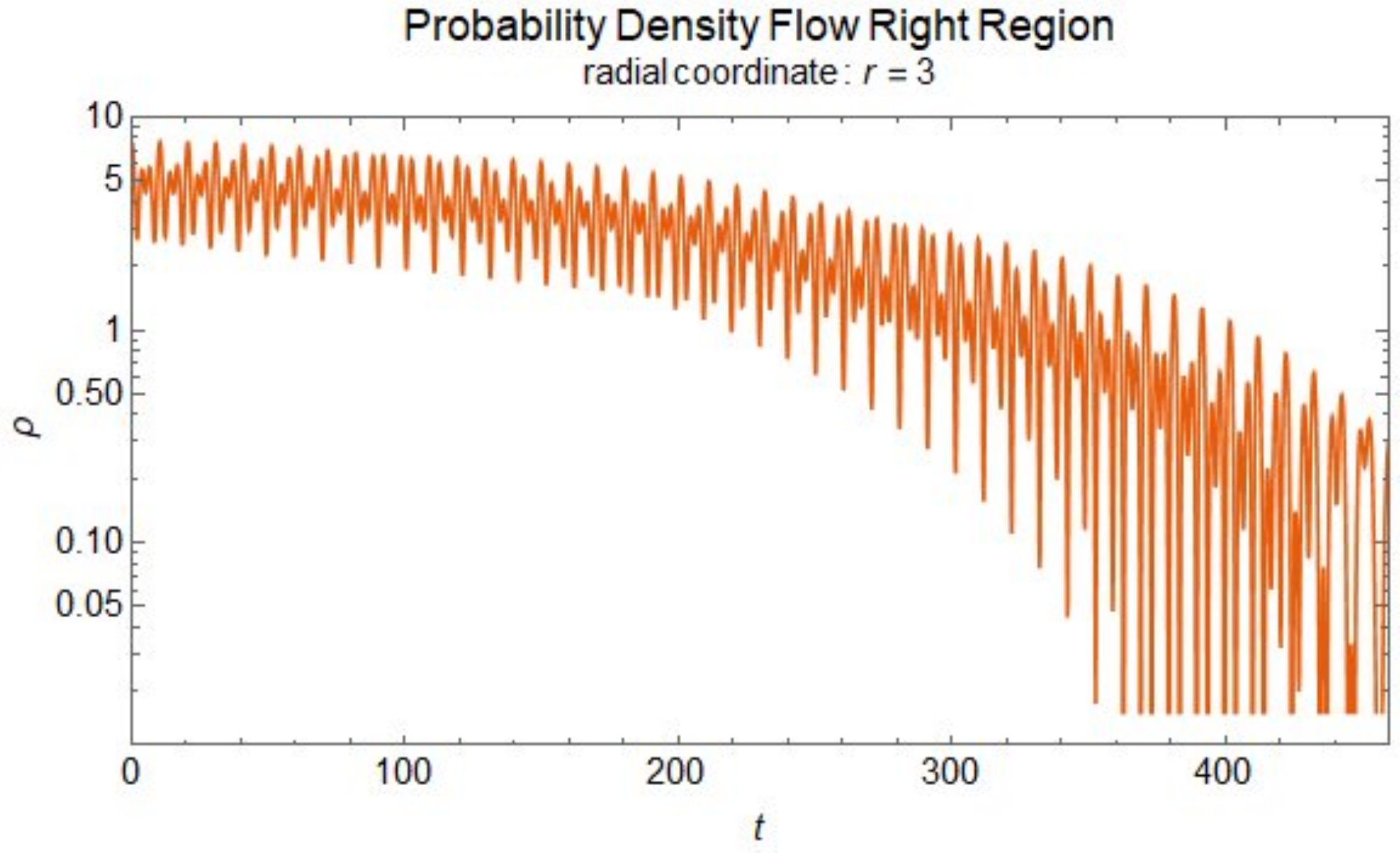}	
				\caption{Flow in the left  region (left panel) for a scalar field in a superposition of $l=0$ and $l=5.$ The time range is equal to half a period. The right panel depicts the corresponding behaviour for the right hand region.}\label{RU05}
			\end{figure}
		\end{center}

	It is of interest to see what will happen if one replaces the contributions of $l=2$ by contributions with $l=0,$ which allow completely free passage between the two regions. Figure \ref{RU05} show us that the $l=0$ contributions just create more  rapidly oscillating behaviour in the flow, while the gross features of the graph still depend on the leading term of $l=5$.

	\section{Extreme mass limit}
\label {sect5}

	 	In our work so far, we have fixed the parameters of the wormhole solution to some fixed values. One would naturally ask if our results are sensitive to the change of these parameters. Taking into account equation (\Ref{M}), we consider the ratio  $a/l_{\eta}$. We observe that when $a$, i.e. the radius of the wormhole, is significantly larger than the non-minimal coupling, the potential peak in the throat seems to vanish.  This means that if the wormhole has a large enough radius, the equation of motion from a test scalar field will not depend on the throat characteristics, rather the test field will just propagate in the background geometry. This observation is made clear from Fig. \ref{flattened}, where we show the drastic change in the potential by changing the throat radius. As the throat radius becomes larger, the Regge-Wheeler potential well becomes wider and the spectrum of the energies tends to a continuum, rendering the corresponding effects unimportant. This is true, even for $m\rightarrow  M_{crit}$, where $M_{crit}$ is given by (\Ref{M}). On the contrary, when $O(a)\leq O(l_{\eta})$, one finds that the throat creates a potential peak, similar to the one we've been working on. A rather interesting result however, is that when $O(a)\leq O(l_{\eta})$ and the mass limits to the critical value, there occurs a qualitative change in the form of the Regge-Wheeler potential, as shown in Fig. \ref{potextr}.
	 	
\begin{center}
 			\begin{figure}[h!]
 				\includegraphics[width=100mm,scale=0.5]{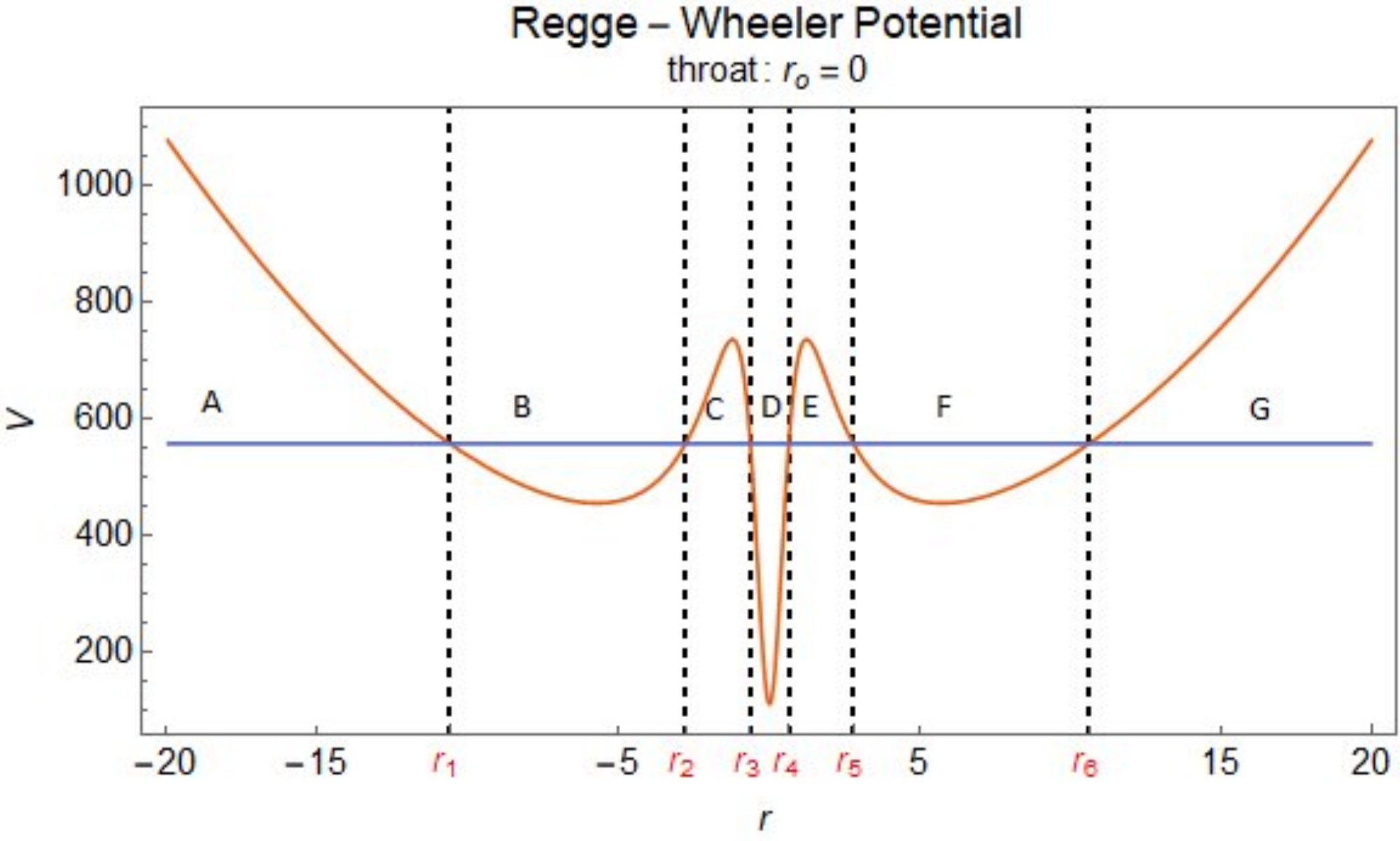}
 				\caption{Regions for the calculation of wave functions.} \label{fig7}
 			\end{figure}
 		\end{center}

\begin{center}
	 		\begin{figure}[h!]
	 			\includegraphics[width=100mm,scale=0.5]{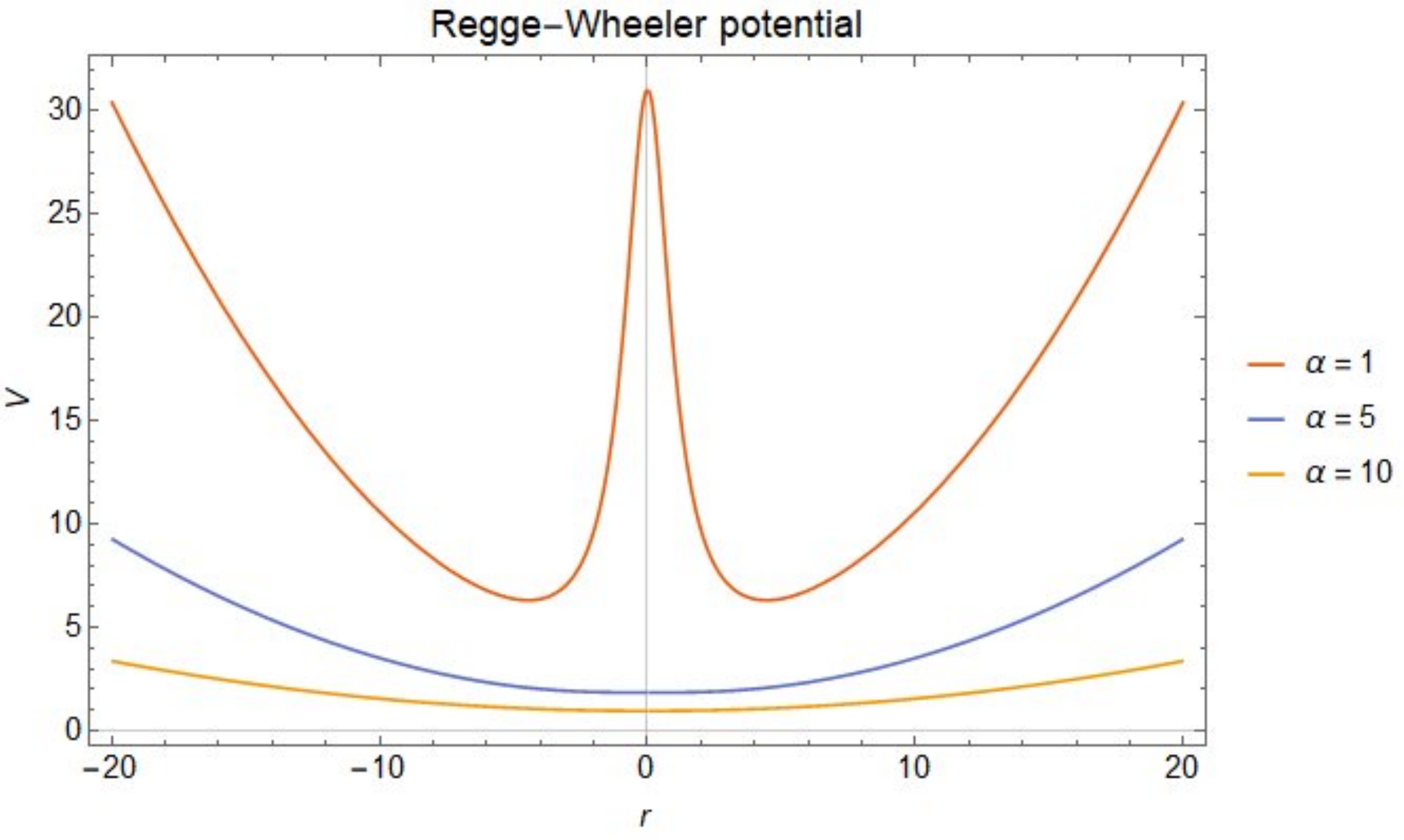}
				\caption{The Regge-Wheeler potential for different values of the radius to coupling ratio, $\alpha$. We have fixed the other parameters to $m=0.1$, $l=5$ and $l_{\eta}=1$}. \label{flattened}
			\end{figure}
	 	\end{center}

\begin{center}
	 	\begin{figure}[h!]
	 		\includegraphics[width=100mm,scale=0.5]{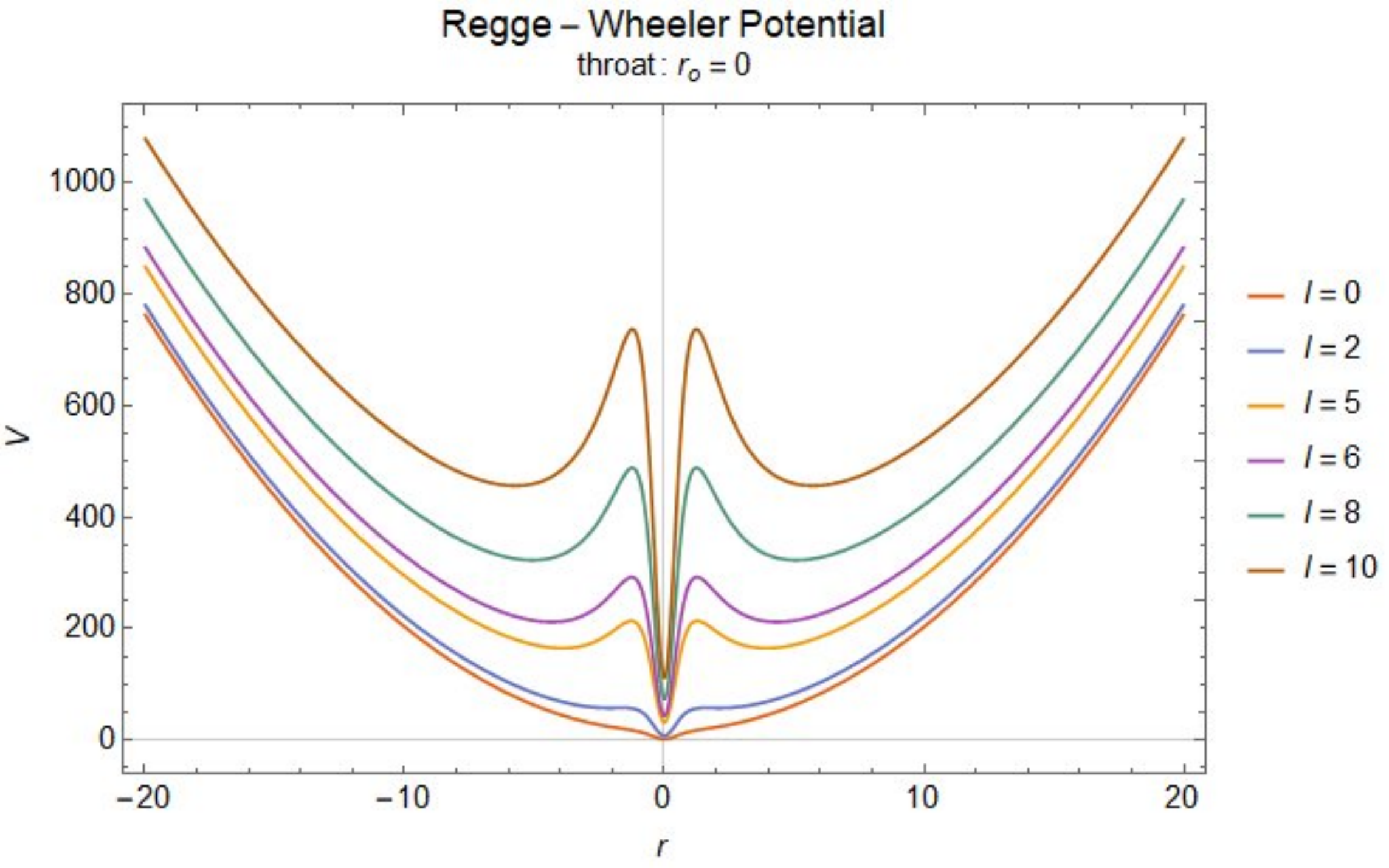}
	 		\caption{Regge-Wheeler Potential for different values of the orbital quantum number. The throat is located at $r=0$. We have fixed the other parameters to $m=0.5,a=1,l_\eta=1$}.\label{potextr}
	 	\end{figure}
	 \end{center}

 			As one can see from Fig. \ref{potextr}, in the limit of  extreme mass $(m\simeq M_{crit}),$ and large orbital quantum number $l,$ a potential well is created at the center, which means that  the lowest energy eigenstates are confined in the throat and cannot escape. This leads to the conclusion that if the test field is in a high orbital state, only sufficiently high energy modes can escape to either region. Note that s-waves can still traverse the wormhole freely. In the following we present the WKB approximation for the above potential for a test field of $l>5$. Similarly to the prior investigation, we will be referring to the global minimum of the potential as $U_{min}$, while the local minima/maxima in each region will be referred to as $V_{min}$/$V_{max}$. If we consider a frequency value of the scalar field, $w^2\in (U_{min},V_{min})$, this energy level cuts the potential twice and the problem reduces to the treatment of a simple potential well problem. As we explain at the end of this section, we have found a couple of bound states in this energy range. For higher energy levels  (yet still smaller than $V_{max}),$ we are going to have seven distinct regions, where the WKB solution changes branch, labeled from left to right as $A$ to $G$, as depicted in Fig. \ref{fig7}. Finally there exist the bound states with energy larger than $V_{max}$ which present no particularly interesting characteristics.
 		 In the following, we present the WKB solutions in each region in the $w^2\in (V_{min},V_{max})$ regime.
 		The solutions in the various regions read
 	\begin{subequations}
 			\begin{align}
 			&U_A(r^*)=\frac{N_L}{\sqrt{|p(r^*)|}}\exp\left[-\int^{r^*_1}_{r^*}|p(r^{*'})|dr^{*'}\right]~,\\
 			&U_B(r^*)=\frac{2N_L}{\sqrt{p(r^*)}}\sin\left(\int^{r^*}_{r^*_1}p(r^{*'})dr^{*'}+\pi/4\right)~,\\
 			&U_C(r^*)=\frac{2N_L \cos\xi}{\sqrt{|p(r^*)|}}\exp\left[\int^{r^*_2}_{r^*}|p(r^{*'})|dr^{*'}\right]+\frac{N_L\sin\xi}{\sqrt{|p(r^*)|}}\exp\left[-\int^{r^*_2}_{r^*}|p(r^{*'})|dr^{*'}\right]~,\\\nonumber
 				\label{emsolution}
 			&U_D(r^*)=\frac{4N_L\cos{\xi} e^{\zeta}}{\sqrt{p(r^*)}}\cos\left[\int^{r^*}_{r^*_3}p(r^{*'})dr^{*'}-\frac{\pi}{4}\right]+\frac{N_L\sin{\xi}e^{-\zeta}}{\sqrt{p(r^*)}}\cos\left[\int^{r^*}_{r^*_3}p(r^{*'})dr^{*'}+\frac{\pi}{4}\right]\\
 			&\quad\quad\quad=\frac{4N_R\cos{\xi} e^{\zeta}}{\sqrt{p(r^*)}}\cos\left[\int_{r^*}^{r^*_4}p(r^{*'})dr^{*'}-\frac{\pi}{4}\right]+\frac{N_R\sin{\xi}e^{-\zeta}}{\sqrt{p(r^*)}}\cos\left[\int_{r^*}^{r^*_4}p(r^{*'})dr^{*'}+\frac{\pi}{4}\right]~,\\
 			&U_E(r^*)=\frac{2N_R \cos\xi}{\sqrt{|p(r^*)|}}\exp\left[\int^{r^*_5}_{r^*}|p(r^{*'})|dr^{*'}\right]+\frac{N_R\sin\xi}{\sqrt{|p(r^*)|}}\exp\left[-\int^{r^*_5}_{r^*}|p(r^{*'})|dr^{*'}\right]~,\\
 			&U_F(r^*)=\frac{2N_R}{\sqrt{p(r^*)}}\sin\left(\int^{r^*_6}_{r^*}p(r^{*'})dr^{*'}+\pi/4\right)~,\\
 			&U_G(r^*)=\frac{N_R}{\sqrt{|p(r^*)|}}\exp\left[-\int^{r^*}_{r^*_6}|p(r^{*'})|dr^{*'}\right]~,
 			\end{align}
 		\end{subequations}
 	where $N_L$ is the amplitude of the field in the left region while $N_R$ is the amplitude of the field in the right region. The different factors found in the solutions are defined as follows:
 	\begin{eqnarray}
 	\label{emdefs}
 	p(r^*)&=&\sqrt{w^2-V^2_{RW}(r^*)}~,\\
 	|p(r^*)|&=&\sqrt{V^2_{RW}(r^*)-w^2}~,\\
 	\xi&=&\int_{r_5^*}^{r_6^*} p(r^{*'})dr^{*'}~,\\
 	\z&=&\int_{r^*_4}^{r^*_5}|p(r^{*'})|dr^{*'}~.
 	\end{eqnarray}
 	The corresponding scattering matrix takes the form,
 		\begin{equation}
 		\label{emsmat}
 		\begin{pmatrix}
 		U_{B_{in}} \\ U_{B_{out}}
 		\end{pmatrix}
 		=
 		\begin{pmatrix}
 		(2e^{2\zeta} + \frac{1}{8}e^{-2\zeta})\cos\chi-i\sin\chi & i(2e^{2\zeta} - \frac{1}{8}e^{-2\zeta})\cos\chi\\
 	-i(2e^{2\zeta} - \frac{1}{8}e^{-2\zeta})\cos\chi & (2e^{2\zeta} + \frac{1}{8}e^{-2\zeta})\cos\chi+i\sin\chi
 		\end{pmatrix}
 		\begin{pmatrix}
 		U_{F_{out}} \\ U_{F_{in}}
 		\end{pmatrix}\\
 		=S\begin{pmatrix}
 		U_{F_{out}} \\ U_{F_{in}}
 		\end{pmatrix}~,\\
 		\end{equation}
 		where the variable $\chi$ is defined as
 		\begin{equation}
 			\chi=\int_{r_3^*}^{r_4^*} p(r^{*'})dr^{*'}~.
 		\end{equation}
 		The quantization condition, similarly to the prior procedure, can be derived from the condition
 		\begin{equation}\label{smatem}
 		S_{11}e^{-2i\xi}i+S_{12}=-iS_{22}e^{2i\xi}+S_{21}~,
 		\end{equation}
 		which yields
 		\begin{equation}
 		\label{qcond}
 		2e^{2\zeta}\cos\chi\cos^2\xi-\frac{1}{8}e^{-2\zeta}\cos\chi\sin^2\xi=\sin\chi\sin\xi\cos\xi~,
 		\end{equation}
 		while the transmission amplitude is given by
 		\begin{equation}\label{Transem}
 		T=\frac{1}{|S_{11}|^2}=\frac{4}{(4e^{2\zeta}+\frac{1}{4}e^{-2\zeta})^2\cos^2\chi+4\sin^2\chi}~.
 		\end{equation}

	 		 	\vspace{-5mm}
 	Equation (\Ref{qcond}) is complicated and difficult to produce intuitively clear results. It is useful to consider even and odd eigenfunctions separately, so that their investigation may be simpler. One can perform an analysis of (\Ref{emsolution}) in terms of odd and even scalar field contribution. We note the alternative forms that follow
 		\be U_D(r^*)&=&N_L\frac{K_1 \cos\frac{\chi}{2}+K_2\sin\frac{\chi}{2}}{\sqrt{p(r^*)}}\cos\left[\int^{r^*}_{0}p(r^{*'})dr^{*'}\right]-N_L\frac{K_1\sin\frac{\chi}{2}-K_2 \cos\frac{\chi}{2}}{\sqrt{p(r^*)}}\sin\left[\int^{r^*}_{0}p(r^{*'})dr^{*'}\right]~,\label{solsleft}\ee 		
 		\be U_D(r^*)&=&N_R\frac{K_1\cos\frac{\chi}{2}+K_2\sin\frac{\chi}{2}}{\sqrt{p(r^*)}}\cos\left[\int^{r^*}_{0}p(r^{*'})dr^{*'}\right]+N_R\frac{K_1 \sin\frac{\chi}{2}-K_2\cos\frac{\chi}{2}}{\sqrt{p(r^*)}}\sin\left[\int^{r^*}_{0}p(r^{*'})dr^{*'}\right]~,\label{solsright}\ee
 where		
 		\be K_1\equiv 4\cos{\xi} e^{\zeta}+\sin\xi e^{-\zeta},\ \ K_2\equiv 4\cos{\xi} e^{\zeta}-\sin\xi e^{-\zeta}~.\ee
 		In what follows we denote the even and odd contributions to the scalar field as $E$ and $O$ respectively. As such,
 		\begin{eqnarray}
 		\label{cont}
	 		E&=&K_1\cos{\frac{\chi}{2}}+K_2\sin{\frac{\chi}{2}}~,\\
	 		O&=&K_1\sin{\frac{\chi}{2}}-K_2\cos{\frac{\chi}{2}}~.
 		\end{eqnarray}
 		 Note that the product of the two contributions, $E$ and $O$, yields the quantization condition, (\Ref{qcond}), as expected.
 		
\begin{center}
 		\begin{figure}[h!]
 			\includegraphics[width=80mm,scale=0.1]{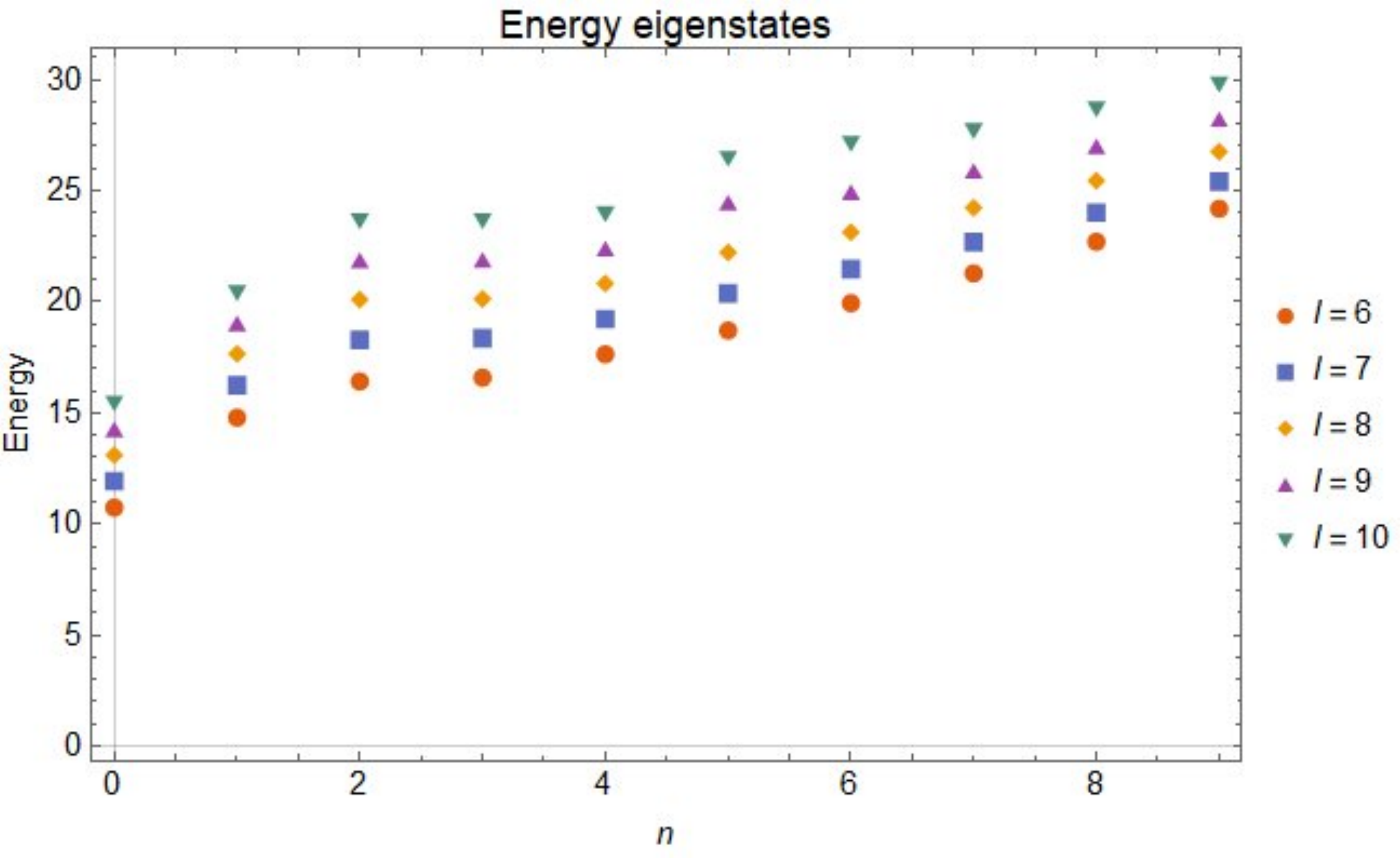}
 			\includegraphics[width=80mm,scale=0.12]{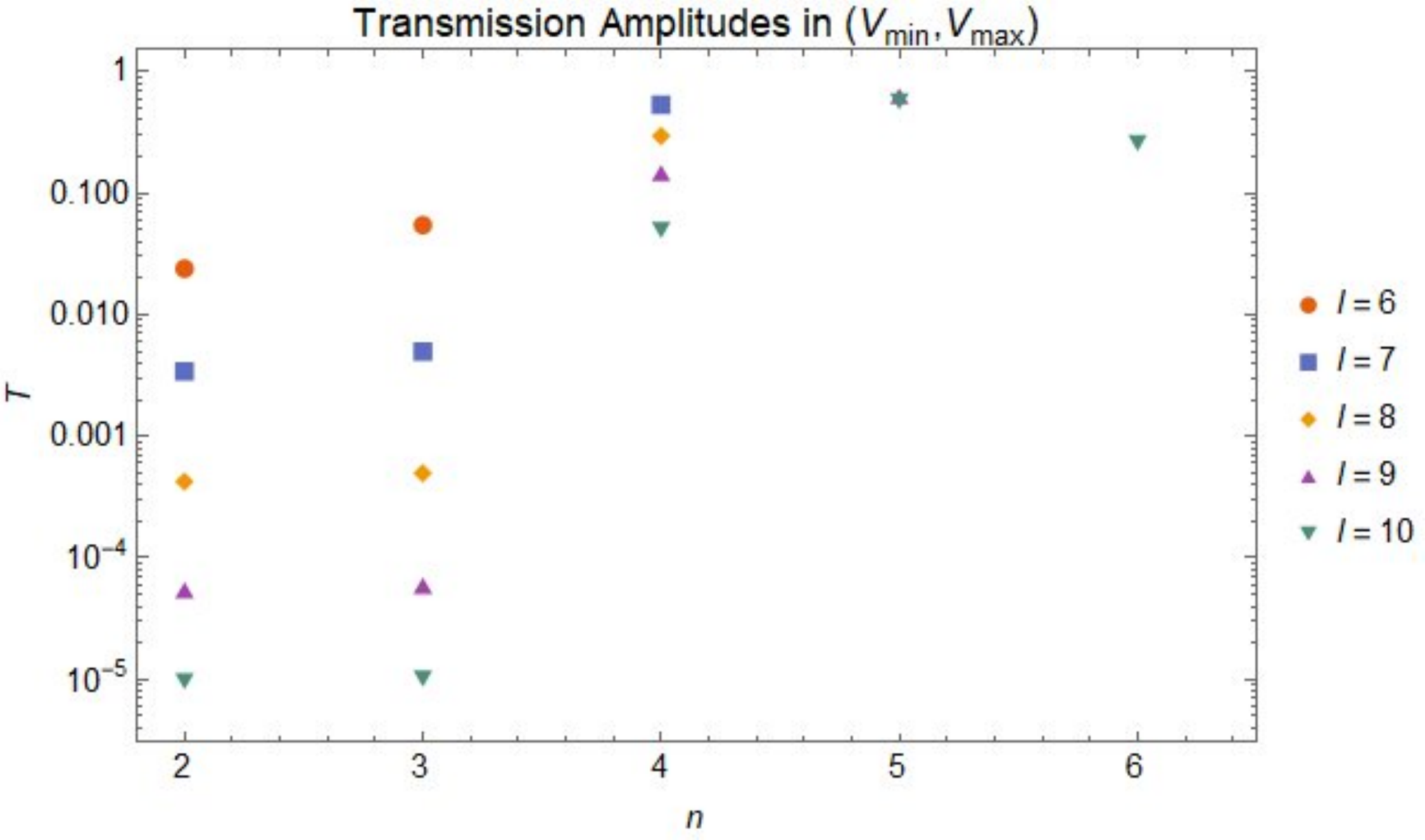}
 			\caption{(a) Ten lowest energy levels in the Regge-Wheeler potential for the values $l=6,\ 7,\ 8,\ 9,\ 10$ of the orbital quantum number $l.$ For $l=10$ one gets the highest energy levels for each value of $n,$ while the lowest ones appear for $l=6.$ (b) Transmission amplitudes for the energy eigenstates in the $(V_{min},V_{max})$ region to pass the throat for various values of the orbital quantum number. The orbital quantum number $l=6$ has the maximum transmission amplitudes for each $n,$ while $l=10$ corresponds to the minimal ones.} \label{entransex}
 		\end{figure}
 	\end{center}

 Therefore, if the scalar field is odd, then the first contribution of (\Ref{solsleft}) and (\Ref{solsright}) is zero and $N_L=-N_R$. Similarly, $N_L=N_R$ if the scalar field is even. Let us consider that the scalar field is even. Then,
 		
 		\begin{equation}
 			O=(4\cos{\xi} e^{\zeta}+\sin\xi e^{-\zeta})\sin\frac{\chi}{2}-(4\cos{\xi} e^{\zeta}-\sin\xi e^{-\zeta})\cos\frac{\chi}{2}=0~,
 		\end{equation}
 		which holds for the solutions
 		\begin{eqnarray}
 		\label{evenfieldquan1}
	 		&&\xi=(n+1)\pi \quad \& \quad \chi=2m\pi+\frac{\pi}{2}~,\\
	 	\label{evenfieldquan2}
	 		&&\xi=(n+\frac{1}{2})\pi \quad \& \quad \chi=2m\pi+\frac{3\pi}{2}~,\\
	 	\label{evenfieldquan3}
	 		&&4 e^{2\zeta}=\tan\xi \frac{\cos\frac{\chi}{2}+\sin\frac{\chi}{2}}{\cos\frac{\chi}{2}-\sin\frac{\chi}{2}}~,
 		\end{eqnarray}
 		where $n,m=0,1,2,..$.
 		Equations (\Ref{evenfieldquan1}-\Ref{evenfieldquan3}) represent the scalar field quantization conditions derived from (\Ref{qcond}) for the case where the scalar field is even. Let us note that, under (\Ref{evenfieldquan1}), the scalar field solution in the throat region, (\Ref{solsleft}), yields $U_D(r^*)\sim e^\zeta$, which is relatively a large number. Therefore, if the transmission amplitude from the throat to either region, $T=e^{-2\zeta}$, is small, we expect the density in these frequencies to be highly concentrated in the throat region. As expected, (\Ref{evenfieldquan1})  maximizes the transmission amplitude (\Ref{Transem}). Thus, even though the density will be highly concentrated in the throat region, the field can still tunnel through from the left to the right region and vice versa.  Similarly, for the odd scalar field case, we find
 		
 		\begin{eqnarray}
 		\label{oddfieldquan1}
 		&&\xi=(n+1)\pi \quad \& \quad \chi=2m\pi+\frac{3\pi}{2}~,\\
 		\label{oddfieldquan2}
 		&&\xi=(n+\frac{1}{2})\pi \quad \& \quad \chi=2m\pi+\frac{\pi}{2}~,\\
 		\label{oddfieldquan3}
 		&&4 e^{2\zeta}=\tan\xi \frac{\sin\frac{\chi}{2}-\cos\frac{\chi}{2}}{\sin\frac{\chi}{2}+\cos \frac{\chi}{2}}~.
 		\end{eqnarray}
 		Our results show that no solution of (\ref{evenfieldquan1})-(\ref{evenfieldquan2}) or (\ref{oddfieldquan1})-(\ref{oddfieldquan2}), which we call resonant solutions, lie in the region $(V_{min},V_{max})$. We found only non-resonant energies,  solutions of (\ref{evenfieldquan3}) and (\ref{oddfieldquan3}). This is because the potential wells created for $l\in[6,10]$ are not big enough to support such states. Thus there remain just the solutions of equations (\ref{evenfieldquan3}) and (\ref{oddfieldquan3}). In the left panel of Fig. (\ref{entransex}) we present our results for all the energy eigenstates up to the ninth excited state. The right panel shows the corresponding transmission amplitudes of the energy eigenstates in the $(V_{min},V_{max})$ region in a logarithmic plot.
 	
The first two energy levels in the left panel of Fig. \ref{entransex} correspond to energy eigenstates trapped in the middle potential well of the throat. This is the reason why the right panel of Fig. \ref{entransex} starts from $n=2$. There occurs an interesting phenomenon in the pattern of the next excited energies. From the left panel of of Fig. \ref{entransex}, one would assume that these energy states are quasi-degenerate. However, the right panel shows that this is not exactly the case, as the $n=4$ energy eigenstate corresponds to a transmission amplitude several order of magnitude higher than the $n=2$ and $n=3$ solution. To explain this result, let us consider the quantization condition (\ref{qcond})
 	\begin{equation}
 	2e^{2\zeta}\cos\chi\cos^2\xi-\frac{1}{8}e^{-2\zeta}\cos\chi\sin^2\xi=\sin\chi\sin\xi\cos\xi~.
 	\end{equation}
 	Since we did not find any resonant energies, we can safely assume that $\xi\neq n\pi/2$ and $\cos(\chi)\neq0$ and divide the expression by $\cos(\chi)\sin^2(\xi)$. The first eigenstates we find in $(V_{min},V_{max})$, i.e. $n=2,3,4$ of the left panel of Fig. \ref{entransex}, correspond to the regime of $e^{2\zeta}>>1$. As such, the $e^{-2\zeta}$  term contributes little and the quantization condition (\ref{qcond}) can be rewritten in a much simpler way as follows
 	\begin{equation}
 	\label{qcondlow}
 	\cot\chi\cot\xi=\frac{1}{2}e^{-2\zeta}~.
 	\end{equation}
 	Performing a qualitative analysis on (\ref{qcondlow}), we find that if $\cot\chi$ takes small values, then this means that we have an energy eigenstate close to a maximum of the transmission amplitude, (\ref{Transem}), but not quite, as is the case of $n=4$ in Fig. \ref{entransex}.
 	If $\cot\chi$ is not close to zero, then this would imply that $\tan\chi$ is just a positive or negative number, but not a particularly large number. Let us set $\tan\chi=\pm c$. As such, equation (\ref{qcondlow}) will yield
 \begin{equation}
 \label{qcondlow1}
\cot\xi=\pm\frac{c}{2}e^{-2\zeta}~,
\end{equation}
 	which is a modification of the previous quantization condition, (\ref{quan}). The $n=2$ and $n=3$ energies can be approximated by (\ref{qcondlow1}). Therefore, there are still corrections enforced by the wormhole geometry, but the quasi-degeneracy is not as pronounced as the previous case, i.e. the lower wormhole mass regime.

The flows we found are of no additional interest and their forms are similar to the ones in figures 5 and 6, albeit with stronger oscillating behaviour due to the high frequency difference of the cross terms that arise due to (\Ref{prob}).
\begin{center}
	\begin{figure}[h!]
		\includegraphics[width=75mm,scale=0.5]{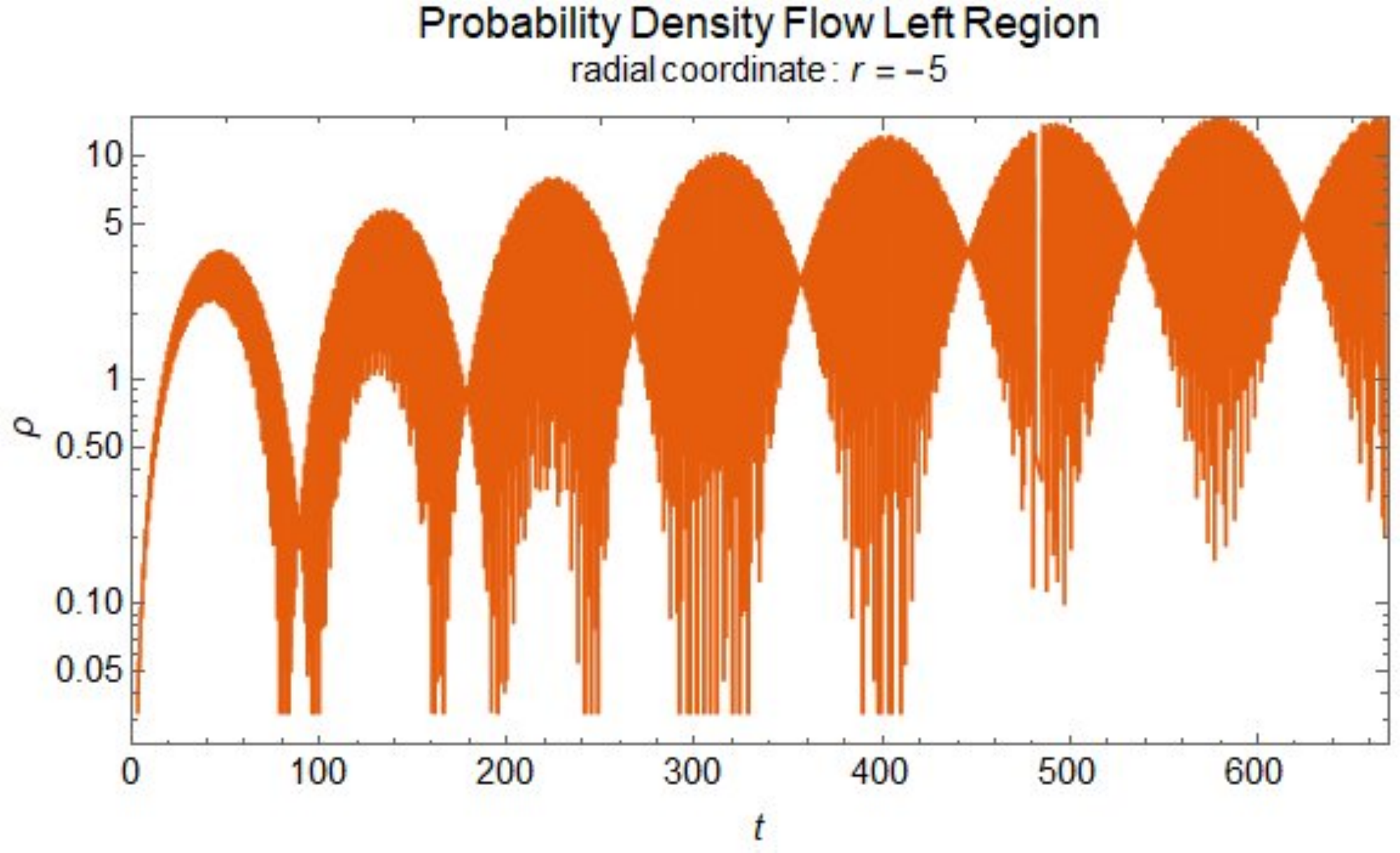}
		\includegraphics[width=75mm,scale=0.5]{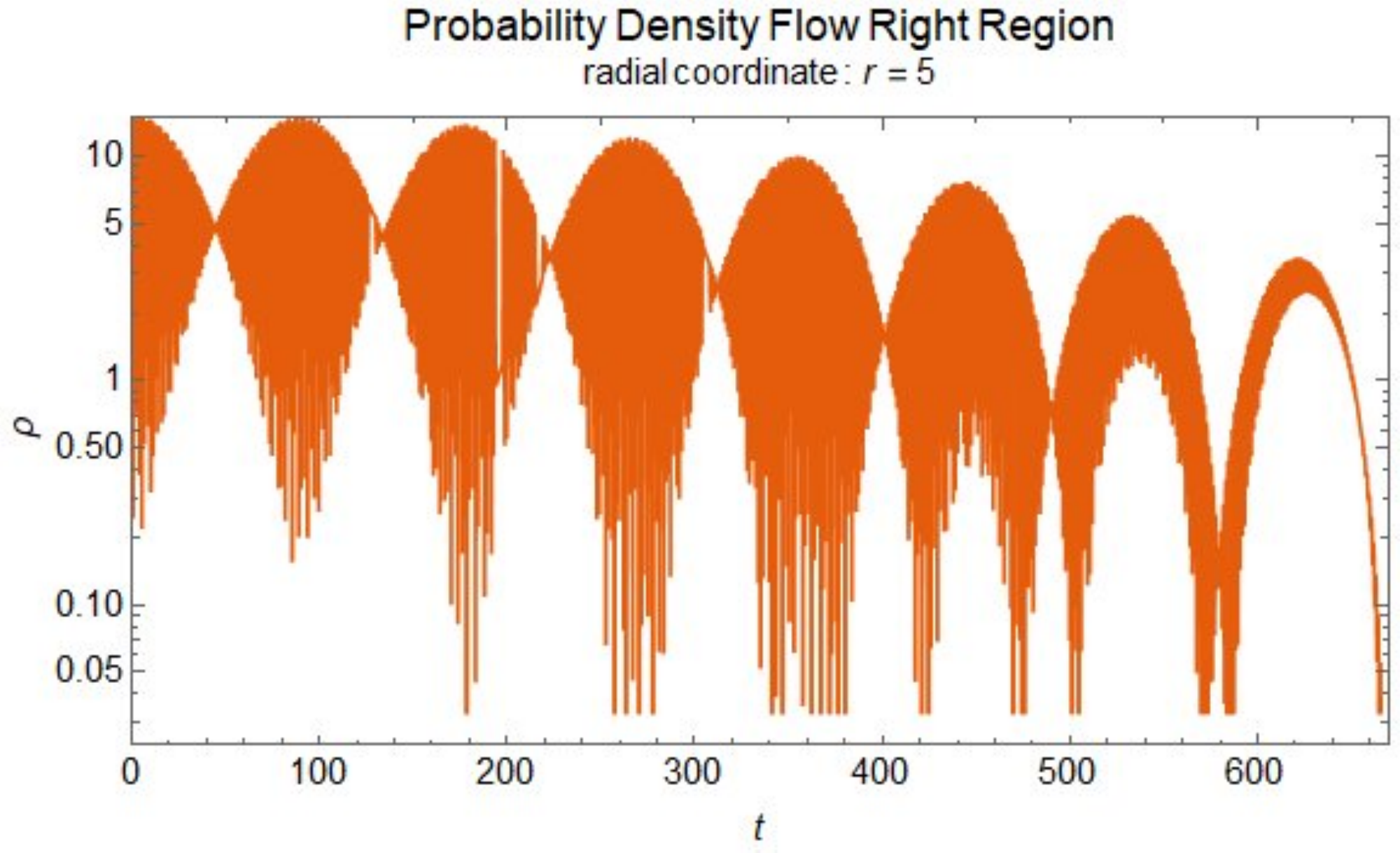}
		\caption{Flow in the left (right)  region for a scalar field in a superposition of $l=7$ and $l=10.$ Each orbital quantum number lies in a superposition of the fundamental and first excited eigenstate in the $(V_{min},V_{max})$ region, to achieve initial localization in the right region. The time range is equal to half a period.}
	\end{figure}
\end{center}

As a consequence of the results, we presume that raising the orbital state of the scalar field or fine-tuning the wormhole characteristics, one would be able to find resonant states. This result according to the prior analysis, would result in a field that is highly localized in the throat region. The maxima of the transmission amplitude, (\Ref{Transem}), can not provide these resonant energies, since it does not account for the right and left potential wells of Fig. (\ref{potextr}). A search for the conditions that allow the existence of resonant energies requires further investigation and we leave it for a future  work.

An interesting issue is that, for this choice of parameters, we have found a couple of low lying states, localized within the well of absolute minimum. These states correspond to the lowest possible energies and, if a particle lies in these states, it is confined in the throat region and cannot escape to either of the two regions.

\section{Conclusions}
\label{sect6}

	We have studied the formation and propagation of bound states in the vicinity of a wormhole when it is described by a non-minimal derivative coupling theory in a gravity theory. The wormhole throat connects two Anti-de Sitter spacetimes. We show that in high orbital states for the scalar field, the corresponding potential has potential barriers that block the passage of a classical particle. We studied the flow from the right-hand region towards the left-hand region for localized states in a single orbital state, as well as in states of superposition of a low and a high value of $l.$ We found out that the flow is greatly dominated by the larger value of $l,$ while the lower value merely induces high frequency oscillations, which perturb the basic oscillation pattern.
	
	It is interesting that, in the limit of large orbital number states and for mass near a critical value, a qualitative change takes place in the potential. In the extreme mass limit, we found that there occurs a creation of a potential well in the throat of the wormhole. This potential well presents a global minimum of the Regge-Wheeler potential. As such, in this regime, the wormhole can trap a low energy scalar particle in the throat region. It is important to note however, that these scalar particles are still accessible from the outside world, due to the lack of an event horizon. For the higher energy eigenstates, we found only non-resonant solutions for the values of the wormhole characteristics we used due to the fact that the potential wells cannot support resonant energies.
 
 If the energy eigenstate is non-resonant, then it corresponds to a low transmission amplitude and the results share several features with the previous case. However, if the energy eigenstate is a resonant solution, then the transmission amplitude is maximized. This implies that the scalar field can tunnel from the first region through the middle potential well and appear in the second region. Due to the reflecting AdS barriers however, the field will become highly localized in the throat. This behaviour is strictly dependent on the existence of the middle potential well and cannot be witnessed in the lower mass case. It will be interesting to test whether these results are shared with other solutions of AdS-asymptotic wormholes.
 
 An extension of this work would be to study what are the effects of the behaviour we found of scalar particles propagating through the wormhole configuration and penetrating the wormhole throat, on early universe cosmology in the context of baby-Universes. To yield some information about the physics of closed Universes, wormholes in AdS spacetimes where discussed in \cite{Maldacena:2004rf}.  Such discussion is connected with the physics of inflation, and its connection with vacuum decay. A very interesting realization of such ideas is the baby-Universe formation by quantum tunneling which eventually is disconnected from the parent spacetime \cite{Giddings:1988wv}. These ideas were further discussed in \cite{Marolf:2020xie} where the negative cosmological constant and the asymptotically AdS boundaries present in the wormhole spacetimes, were connected to baby-Universes.

\section{Appendix: S matrix and quantization condition}

We begin with the Klein Gordon equation
\begin{equation}
\frac{d^2 u}{dr^{* 2}}+(w^2-V_{eff}(r^*))u=0~. \label{1}
\end{equation}
At each turning point, $r^*_i$, we perform a first order Taylor expansion of the form
\begin{equation}
	p^2(r^*)=w^2-V_{eff}(r^*)\approx-V_{eff}^\prime(r^*_{i})(r^*-r^*_i)=-\lambda_i (\pm)(r^*-r^*_i)~, \label{2}
\end{equation}
where the $\pm$ signature is needed to keep all the $\lambda_i$ positive.
Then, perform a change of variables of the following form
\begin{equation}
y=\lambda_i^{1/3} (\pm)(r^*-r^*_i)=-\lambda_i^{-2/3}p^2(r^*)~. \label{3}
\end{equation}
Under (\ref{3}), the differential equation (\ref{1}) takes the following form
\begin{equation}
\left[\frac{d^2}{dy^2}-y\right]U(y)=0~. \label{4}
\end{equation}
This equation has known solutions which are the Airy functions.

Equation (\ref{4}) has a solution of the form: $U(y)=M_1\textbf{Ai}(y)+M_2\textbf{Bi}(y)$. We are going to use the asymptotic formulas of the Airy functions.
\begin{enumerate}
	\item
	At $y<<0$, we have
	\begin{equation}
	\begin{split}
	U(y)\sim\frac{M_1e^{-i\pi/4}}{2\sqrt{\pi}\abs{y}^{1/4}}\exp(i\frac{2}{3}\abs{y}^{3/2})
+\frac{M_1e^{i\pi/4}}{2\sqrt{\pi}\abs{y}^{1/4}}\exp(-i\frac{2}{3}\abs{y}^{3/2})
+\\\frac{M_2e^{i\pi/4}}{2\sqrt{\pi}\abs{y}^{1/4}}\exp(i\frac{2}{3}\abs{y}^{3/2})
+\frac{M_2e^{-i\pi/4}}{2\sqrt{\pi}\abs{y}^{1/4}}\exp(-i\frac{2}{3}\abs{y}^{3/2})~.
	\end{split} \label{5}
	\end{equation}
	\item
	At $y>>0$, we have
	\begin{equation}
	U(y)\sim\frac{M_1}{2\sqrt{\pi}y^{1/4}}\exp(-\frac{2}{3}y^{3/2})+\frac{M_2}{\sqrt{\pi}y^{1/4}}\exp(\frac{2}{3}y^{3/2})~. \label{6}
	\end{equation}
\end{enumerate}
This means that for each region, the WKB wavefunctions must be validated by the corresponding Airy solutions.
Using the WKB wavefunctions we found for the first case, we reach the following set of equations for each region of the wormhole
\begin{itemize}
	\item Right Region
\begin{itemize}
	\item Region E
	\begin{equation}
		U_E(r^*)=\frac{E}{\sqrt{|p(r^*)|}}\exp\left[-\int^{r^*}_{r^*_4}|p(r^{*'})|dr^{*'}\right]~.\label{7}\\
	\end{equation}
	The Taylor expansion is done on $r^*_4$ and the corresponding amplitudes of the WKB solutions with respect to the Airy functions read
	\begin{align}
		E=\frac{M_1\lambda_4^{1/6}}{2\sqrt{\pi}}~,\label{8}\\
		M_2=0 \label{9}~.
	\end{align}
		\item Region D
		\begin{enumerate}
			\item Right turning point
	\begin{equation}
	U_D(r^*)=\frac{D_{1DE}}{\sqrt{p(r^*)}}\exp\left[i\int^{r^*_4}_{r^*}p(r^{*'})dr^{*'}\right]+\frac{D_{2DE}}{\sqrt{p(r^*)}}\exp\left[-i\int^{r^*_4}_{r^*}p(r^{*'})dr^{*'}\right]~. \label{10}
	\end{equation}
	The Taylor expansion is done on $r^*_4$ and the corresponding amplitudes of the WKB solutions with respect to the Airy functions read
	\begin{align}
	D_{1DE}=\frac{M_1\lambda_4^{1/6}e^{-i\pi/4}}{2\sqrt{\pi}}~, \label{11}~, \\
	D_{2DE}=\frac{M_1\lambda_4^{1/6}e^{+i\pi/4}}{2\sqrt{\pi}}~ \label{12}~.
	\end{align}
	\item Left turning point
	\begin{equation}
		U_D(r^*)=\frac{D_{1CD}}{\sqrt{p(r^*)}}\exp\left[i\int^{r^*}_{r^*_3}p(r^{*'})dr^{*'}\right]+\frac{D_{2CD}}{\sqrt{p(r^*)}}\exp\left[-i\int^{r^*}_{r^*_3}p(r^{*'})dr^{*'}\right]~. \label{13}
	\end{equation}
	The Taylor expansion is done on $r^*_3$ and the corresponding amplitudes of the WKB solutions with respect to the Airy functions read
		\begin{align}
	D_{1CD} = \lambda_3^{1/6} \left( \frac{L_1 e ^{-i\pi / 4}}{2 \sqrt{\pi}} + \frac{L_2 e ^{i\pi / 4} }{2 \sqrt{\pi}} \right)~, \label{14} \\
	D_{2CD} =\lambda_3^{1/6} \left( \frac{L_1 e ^{i\pi / 4}}{2 \sqrt{\pi}} + \frac{L_2 e ^{-i\pi / 4} }{2 \sqrt{\pi}} \right)~, \label{15}
	\end{align}
	\item Connection of solutions
	\begin{align}
	D_{1CD} \exp\left[ i \xi\right] = D_{2DE}~, \label{16} \\
	D_{2CD} \exp\left[ -i \xi\right] = D_{1DE}~. \label{17}
	\end{align}
	\end{enumerate}
	\item Region C:
	\begin{equation}
	U_C(r^*)=\frac{C_{1CD}}{\sqrt{|p(r^*)|}}\exp\left[\int^{r^*_3}_{r^*}|p(r^{*'})|dr^{*'}\right]+\frac{C_{2CD}}{\sqrt{|p(r^*)|}}\exp\left[-\int^{r^*_3}_{r^*}|p(r^{*'})|dr^{*'}\right]~. \label{18}
	\end{equation}
	The Taylor expansion is done on $r^*_3$ and the corresponding amplitudes of the WKB solutions with respect to the Airy functions read
	\begin{align}
	C_{1CD}=\frac{L_2\lambda_3^{1/6}}{\sqrt{\pi}}~, \label{19} \\
	C_{2CD}=\frac{L_1\lambda_3^{1/6}}{2\sqrt{\pi}}~. \label{20}
	\end{align}	
\end{itemize}
\item Left Region
	\begin{itemize}
		\item Region A
		\begin{equation}
		U_A(r^*)=\frac{A}{\sqrt{|p(r^*)|}}\exp\left[-\int^{r^*_1}_{r^*}|p(r^{*'})|dr^{*'}\right]~.\label{21}\\
		\end{equation}
		The Taylor expansion is done on $r^*_1$ and the corresponding amplitudes of the WKB solutions with respect to the Airy functions read
		\begin{align}
		A=\frac{J_1\lambda_1^{1/6}}{2\sqrt{\pi}}~.\label{22}\\
		J_2=0~. \label{23}
		\end{align}
		\item Region B
		\begin{enumerate}
			\item Left turning point
		\begin{equation}
		U_B(r^*)=\frac{B_{1AB}}{\sqrt{p(r^*)}}\exp\left[i\int^{x*}_{r^*_1}p(r^{*'})dr^{*'}\right]+\frac{B_{2AB}}{\sqrt{p(r^*)}}\exp\left[-i\int^{x*}_{r^*_1}p(r^{*'})dr^{*'}\right]~. \label{24}
		\end{equation}
		The Taylor expansion is done on $r^*_1$ and the corresponding amplitudes of the WKB solutions with respect to the Airy functions read
		\begin{align}
		B_{1AB}=\frac{J_1\lambda_1^{1/6}e^{-i\pi/4}}{2\sqrt{\pi}}~,  \label{25}\\
		B_{2AB}=\frac{J_1\lambda_1^{1/6}e^{+i\pi/4}}{2\sqrt{\pi}}~.  \label{26}
		\end{align}
		\item Right turning point
		\begin{equation}
		U_B(r^*)=\frac{B_{1BC}}{\sqrt{p(r^*)}}\exp\left[i\int_{x*}^{r^*_2}p(r^{*'})dr^{*'}\right]+\frac{B_{2BC}}{\sqrt{p(r^*)}}\exp\left[-i\int_{x*}^{r^*_2}p(r^{*'})dr^{*'}\right]~.  \label{27}
		\end{equation}
		The Taylor expansion is done on $r^*_2$ and the corresponding amplitudes of the WKB solutions with respect to the Airy functions read
		\begin{align}
		B_{1BC} = \lambda_2^{1/6} \left( \frac{K_1 e ^{-i\pi / 4}}{2 \sqrt{\pi}} + \frac{K_2 e ^{i\pi / 4} }{2 \sqrt{\pi}} \right)~,  \label{28} \\
		B_{2BC} =\lambda_2^{1/6} \left( \frac{K_1 e ^{i\pi / 4}}{2 \sqrt{\pi}} + \frac{K_2 e ^{-i\pi / 4} }{2 \sqrt{\pi}} \right)~.  \label{29}
		\end{align}
		\item Connection of solutions
		\begin{align}
			B_{1AB} \exp\left[ i \xi\right] = B_{2BC}~,  \label{30} \\
			B_{2AB} \exp\left[ -i \xi\right] = B_{1BC}~, \label{31}
		\end{align}
		\end{enumerate}
	\item Region C
		\begin{equation}
		U_C(r^*)=\frac{C_{1BC}}{\sqrt{|p(r^*)|}}\exp\left[\int_{r^*_2}^{r^*}|p(r^{*'})|dr^{*'}\right]+\frac{C_{2BC}}{\sqrt{|p(r^*)|}}\exp\left[-\int_{r^*_2}^{r^*}|p(r^{*'})|dr^{*'}\right]~.  \label{32}
		\end{equation}
		The Taylor expansion is done on $r^*_2$ and the corresponding amplitudes of the WKB solutions with respect to the Airy functions read:
		\begin{align}
		C_{1BC}=\frac{K_2\lambda_2^{1/6}}{\sqrt{\pi}}~,   \label{33}\\
		C_{2BC}=\frac{K_1\lambda_2^{1/6}}{2\sqrt{\pi}}~.   \label{34}
		\end{align}	
	\end{itemize}
\end{itemize}
From the above equations, one can deduce the WKB wavefunctions. Note that the connection of solutions for region C is
\begin{align}
C_{1BC}=C_{2CD}e^{-\zeta}~,  \label{35}\\
C_{2BC}=C_{1CD}e^{\zeta}~.  \label{36}
\end{align}
Now for the scattering matrix and the quantization condition, we start grouping stuff together:
\begin{itemize}
	\item The scattering matrix, $S$, is of the form
	\begin{equation}
	\begin{pmatrix}
	B_{2BC}\\B_{1BC}
	\end{pmatrix}
	=S
	\begin{pmatrix}
		D_{1CD}\\D_{2CD}
	\end{pmatrix}~.  \label{37}
	\end{equation}
	To deduce the form of the scattering matrix we can use equations (\ref{28}-(\ref{29}), then plug in (\ref{33}-(\ref{34}), use the connections of (\ref{35}-(\ref{36}), plug in (\ref{19}-(\ref{20}) and finally plug in (\ref{14}-(\ref{15}). That is:
	\begin{eqnarray}
	\begin{pmatrix}
	B_{2BC}\\B_{1BC}
	\end{pmatrix}&=&
	\begin{pmatrix}
	\lambda_2^{1/6} \left( \frac{K_1 e ^{i\pi / 4}}{2 \sqrt{\pi}} + \frac{K_2 e ^{-i\pi / 4} }{2 \sqrt{\pi}} \right)\\\lambda_2^{1/6} \left( \frac{K_1 e ^{-i\pi / 4}}{2 \sqrt{\pi}} + \frac{K_2 e ^{i\pi / 4} }{2 \sqrt{\pi}} \right)
	\end{pmatrix}   \label{38}\\&=&
	\begin{pmatrix}
	C_{2BC}e^{i\pi/4}+\frac{C_{1BC}e^{-i\pi/4}}{2}\\
	C_{2BC}e^{-i\pi/4}+\frac{C_{1BC}e^{i\pi/4}}{2}
	\end{pmatrix}   \label{39}\\&=&
	\begin{pmatrix}
 \frac{L_2\lambda_3^{1/6}}{\sqrt{\pi}}e^\zeta e^{i\pi/4}+\frac{\frac{L_1\lambda_3^{1/6}}{2\sqrt{\pi}}e^{-\zeta}e^{-i\pi/4}}{2}\\
\frac{L_2\lambda_3^{1/6}}{\sqrt{\pi}}e^\zeta e^{-i\pi/4}+\frac{\frac{L_1\lambda_3^{1/6}}{2\sqrt{\pi}}e^{-\zeta}e^{i\pi/4}}{2}
	\end{pmatrix}  \label{40}\\&=&
	\begin{pmatrix}
			e^{\zeta} + \frac{1}{4}e^{-\zeta} & i(e^{\zeta} - \frac{1}{4}e^{-\zeta})\\
		-i (e^{\zeta} - \frac{1}{4}e^{-\zeta}) & e^{\zeta} + \frac{1}{4}e^{-\zeta}
	\end{pmatrix}
	\begin{pmatrix}
			D_{1CD}\\D_{2CD}
	\end{pmatrix}~.   \label{41}
	\end{eqnarray}

	\item For the quantization condition we follow the same procedure starting from (\ref{41}) and "going backwards". That is:
	We use the connections of the solutions (\ref{30}-\ref{31}) and plug in (\ref{25}-\ref{26}) and finally (\ref{22}) for the left hand side. Similarly, we use the connections of the solutions (\ref{16}-\ref{17}) and plug in (\ref{11}-\ref{12}) and finally (\ref{8}) for the right hand side. Therefore, the left hand side of (\ref{41}) will read
	\begin{equation}
		\begin{pmatrix}
		B_{2BC}\\B_{1BC}
		\end{pmatrix}
		=
		\begin{pmatrix}
		e^{i\xi}&0\\0&e^{-i\xi}
		\end{pmatrix}
		\begin{pmatrix}
		Ae^{-i\pi/4}\\Ae^{i\pi/4}
		\end{pmatrix}~.  \label{42}
	\end{equation}
		while the right hand side of (\ref{41}) will read
	\begin{equation}
	\begin{pmatrix}
	D_{1CD}\\D_{2CD}
	\end{pmatrix}
	=
	\begin{pmatrix}
	e^{-i\xi}&0\\0&e^{i\xi}
	\end{pmatrix}
	\begin{pmatrix}
	Ee^{i\pi/4}\\Ee^{-i\pi/4}
	\end{pmatrix}~.  \label{43}
	\end{equation}
	Using now (\ref{41}-\ref{43}), one can find that
	\begin{equation}
	\begin{pmatrix}
	Ae^{-i\pi/4}\\Ae^{i\pi/4}
	\end{pmatrix}=
	\begin{pmatrix}
	e^{-i\xi}&0\\0&e^{i\xi}
	\end{pmatrix}S
	\begin{pmatrix}
	e^{-i\xi}&0\\0&e^{i\xi}
	\end{pmatrix}
	\begin{pmatrix}
	Ee^{i\pi/4}\\Ee^{-i\pi/4}
	\end{pmatrix}~.  \label{44}
	\end{equation}
	The system of equations (\ref{44}) has a non-zero solution if
	\begin{equation}
	S_{11}e^{-2i\xi}i+S_{12}=-iS_{22}e^{2i\xi}+S_{21}~,  \label{45}
	\end{equation}
the solution of which yields the quantization condition.

\end{itemize}

\end{document}